\documentclass[11 pt,a4paper]{article}

\usepackage[T1]{fontenc}
\usepackage[spanish,english]{babel}
\usepackage[space]{grffile}     
\usepackage{amsmath}
\usepackage{amssymb}
\usepackage{mathptmx}
\usepackage{txfonts}

\usepackage{graphicx} 

\usepackage{multicol} 
\usepackage{float}
\usepackage[comma]{natbib}	
\usepackage{multicol}
\usepackage{caption}
\usepackage{float}
\usepackage{color}
\usepackage{rotating}
\usepackage{enumitem}

\usepackage{xcolor}
\usepackage{hyperref}

\usepackage{lipsum}

 \title{Capturing the short-term characteristics of a barred galaxy from a single snapshot}

\author{
P. S\'anchez-Mart\'{i}n$^1$\thanks{patricia.sanchez.martin@upc.edu} \and
J. Amor\'os$^{1,2,3}$
\and J. J. Masdemont$^{1,2,3,4}$
}

\date{
$^1$Dept. de Matem\`{a}tiques, Universitat Polit\`{e}cnica de Catalunya, Barcelona, Spain\\
$^2$Centre de Recerca Matem\`atica, Bellaterra, Spain\\
$^3$IMTEch, Universitat Polit\`{e}cnica de Catalunya, Barcelona, Spain \\
$^4$IEEC, Barcelona, Spain
}

\begin{document}

\providecommand{\keywords}[1]{\textbf{\textit{keywords --}} #1}

\maketitle

\begin{abstract}
Using a dataset of stars from a barred galaxy that provides instantaneous positions and velocities in the sky, we propose a method to identify its key structures, such as the central bar and spiral arms, and to determine the bar pattern speed in the short term.
This is accomplished by applying combinatorial and topological data analysis techniques to the available star-state information set.

To validate the methodology, we apply it to two scenarios: a test particle simulation and a N-body simulation, both evaluated when the bar is fully formed. In both cases, the results are robust and consistent. We then use these outcomes to calibrate a classical dynamical model and compute related structures, such as equilibrium points and invariant manifolds, which we verify align with the galaxy's spiral arms.

\end{abstract}



\section{Introduction}

In recent years, the analysis of the angular velocity of the bar (usually called the pattern speed of the bar) in barred galaxies using a single snapshot has become an active area of research, owing to the availability of accurate astrometric data from surveys such as Gaia for the Milky Way and its neighbourhood~\citep{Gaia2021}. The goal of this work is to develop a method to extract the main characteristics of a barred galaxy (bar pattern speed, size, mass and location of the main galactic components) from positions and velocities of the stars in a unique instant of time, as given e.g. by Gaia. These data are used afterwards to adjust the components of the galaxy in a dynamical system model, which allows the study of the dynamics of the galaxy in a simple way instead of working with millions of constituting stars. 

Barred galaxies are usually modelled as a dynamical system where the different components of the galaxy are introduced as potentials. The main components of these galaxies are a central bar, a disc surrounding the bar and a halo (disc) of dark matter enveloping the bar and disc. The potentials that describe these components can have simple expressions. This facilitates the study of the resulting dynamical system, using series expansions of high orders such as logarithmic potentials~\citep{Rom09}, or more popular complex potentials~\citep[see e.g.][]{Athan1983,Katsanikas2022,Skokos,Tsoutsis2008,Tsoutsis2009}.

\begin{figure}
\begin{center}
\includegraphics[width=0.7\textwidth]{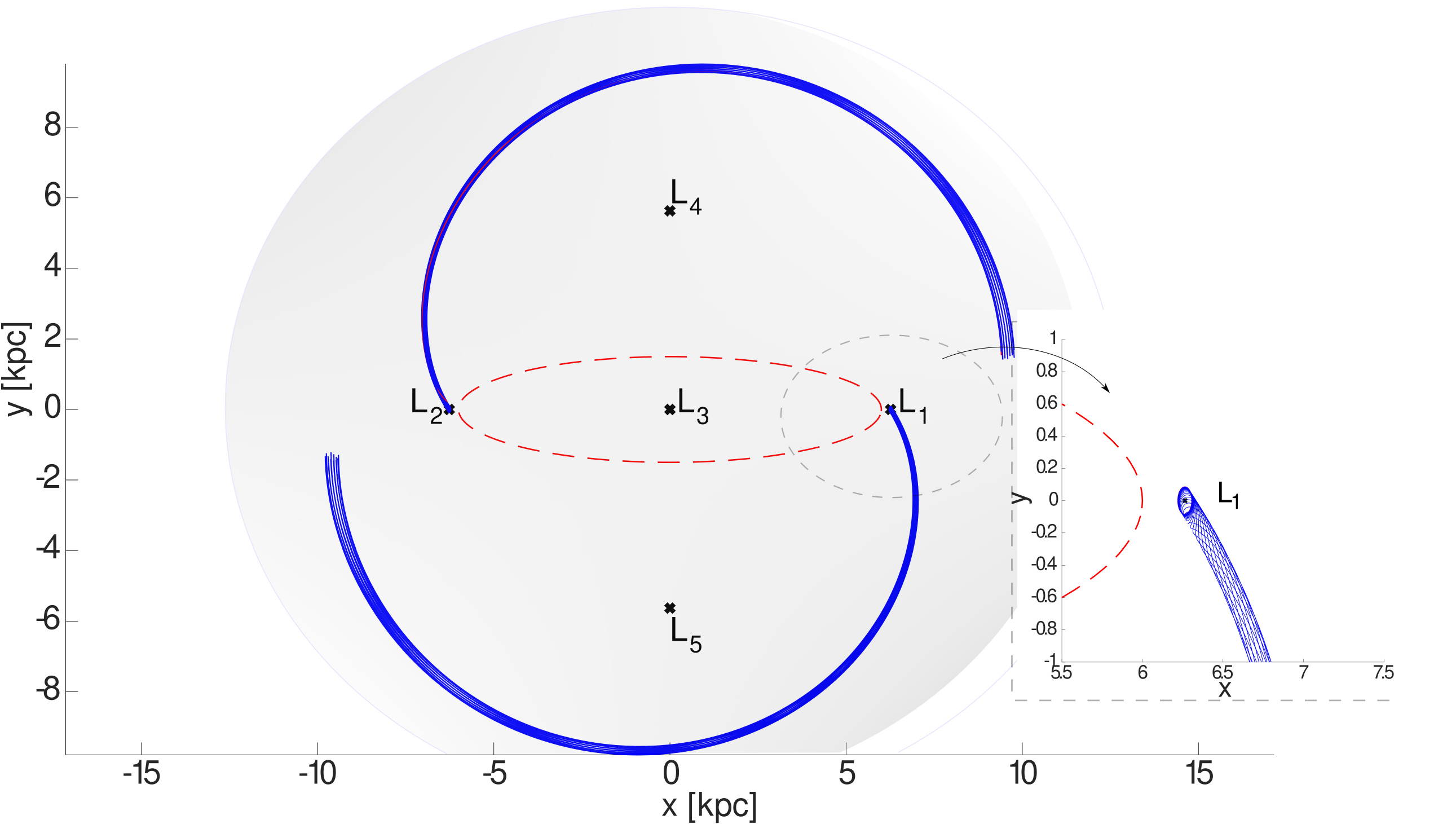}
\end{center}
\caption{Dynamics of a barred galaxy model. The central bar is delineated by a dotted red curve, the disc surrounding the bar is coloured in grey, the halo would be surrounding all the components up to 100 kpc. The Lagrangian points are marked in black. The invariant manifolds that emanate from the Lyapunov periodic orbits are plotted in blue. A zoom around one of the unstable equilibrium points is displayed on the right side.}
\label{fig:galaxy_components}
\end{figure}

This dynamical system model (see Fig.~\ref{fig:galaxy_components}) behaves, to some extent, as the Restricted Three Body Problem. In the rotating (synodical) reference frame, where the $x$-axis is aligned with the main axis of the bar, the model has five Lagrangian equilibrium points for common values of the parameters of the potentials: Two of them placed at each side of the bar end and are of type saddle $\times$ centre $\times$ centre, one in the centre of the bar with a centre $\times$ centre $\times$ centre behaviour and two located up and down the bar with the same behaviour as the centre point. The behaviour of the points placed at the ends of the bar allows the existence of Lyapunov periodic orbits and invariant tori around the equilibrium points, which act as gates between the inner and outer region of the galaxy, delimited by the zero velocity curves. The invariant manifolds that emanate from these periodic orbits and invariant tori support the movement of stars between both regions, manifesting themselves as the arms of galaxies\citep{Rom09,Romero1,Romero2011, Warps, Asymmetry,Tsoutsis2008}. The different shapes of barred galaxies (ring, spiral) is related with the existence of homoclinic or heteroclinic orbits, while the transit orbits contained in the interior of the manifolds are responsible for this transfer of matter~\citep{Romero1,Romero2,GideaMasdem2007,HenryScheeres2024}. These transit orbits are also widely studied in the context of the Restricted Three Body Problem~\citep[see e.g.][]{GomezKoon2004, KoonLo2000, CanaliasMasdemont2006, Fantino2023}. 

A widely used method for determining the pattern speed was introduced by \citet{TW1984}, analysing the surface brightness of stars and assuming that they fulfill the continuity equation~\citep[see e.g.][]{Peschken2019,Fragkoudi2021}. Another common tool is to use the $m=2$ mode of the Fourier expansion to find the position and pattern speed of the bar. These estimates have a sizeable error as they ignore nontrivial contributions such as those of the $m=4,6,8$ modes~\citep{Pfenniger2023}, and to reduce this error it is common to use several snapshots of a simulation to compute the angular velocity of the bar by finite differences~\citep{SellwoodAthan1986}.

To determine the pattern speed of the bar in a single snapshot, \citet{Marchuk2024} searches the stars around the unstable equilibrium points using photometric methods. An alternative approach is followed in \citet{Kalda2024}: Its authors use a deep learning tool to analyze the motion of the observed stars and infer from it a gravitational potential describing the galaxy. In order to do this, fictitious stars have to be added to the galaxy. The resulting potential is complex, unsuitable for analytic study and computationally costly to use. In \citet{Dehnen2023}, its authors propose a method for determining the orientation angle and pattern speed of simulated barred galaxies using single snapshots, expanding the method to include the calculation of the rate and axis of rotation. This approach entails computing the Fourier transform of the surface density of the galaxy, and fitting the resulting pattern speed and orientation angle to the observed morphology. \citet{Pfenniger2023} reviews methods for ascertaining the bar and spiral pattern speeds of galaxies. These methods involve measuring the rotation rates of functions based on the coordinates or velocities of particles within the galaxy and can be used to determine various parameters such as the pattern speed vector, potential pattern speed, mode rotation speeds, and pattern speed accelerations. \citet{JimenezArranz2024} probes the method proposed by \citet{Dehnen2023} and one of the method in \citet{Pfenniger2023} in a simulation of the Large Magellanic Cloud which uses the data given by Gaia, achieving different results with each method. 

In this paper, we propose a methodology for detecting key temporary characteristics of a galaxy, 
using only a single snapshot of the instantaneous position and velocity of its stars obtained either from simulations or as observational data. Our approach does not rely on density regions. Instead,  by combinatorial and topological data analysis techniques,
we identify groups of particles following particular motions. 
No neural network has been necessary for these determinations: our procedure involves applying k-means grouping to star velocities and persistent homology to their positions. The resulting determination of the bar features shows some dependence on the strictness selected in the basic algorithms but responds to deterministic criteria, which can be explicitly described. 
 
The stars are classified as temporarily belonging to the bar, not only based on their positions but also considering their radial velocities, in contrast to disc particles that occupy the same space. We have observed that this distinction significantly reduces errors in the computation of the bar angular velocity. As we have said, at the central part of the galaxy, inside the bar, is placed an equilibrium point which behaves as a linearly stable point. Around this point exists a main family of periodic orbits called the $x_1$ family that, together with several bifurcations of it, gives structure and consistency to the bar~\citep{Athan1983,Patsis2019,Pfenn,Skokos}. We use the dynamics of this family to detect the stars that pertain to the bar in the single snapshot. A similar idea has been used in~\citet{Petersen2024} combining the velocity field with Fourier expansions, calling the bar find in this way the ''dynamical bar''.
 
The procedure enables the identification of the central bar of the galaxy and its features, such as position, length, pattern speed (both instantaneous and short-term average), density distribution, and galactic arms,
in spite of the fact
that particles do not remain bound to these defined structures over time; rather, as we have observed, they continuously move in and out,
emphasizing that the bar is not a solid rigid body and the continuous interchange of disc and bar particles is taken into account to adjust the pattern speed of the bar in the dynamical model over short timescales.
However, the estimates obtained with this method remain consistent, indicating that while individual stars may change roles, the system's overall properties remain stable in the short term.

To find the start of the manifolds (arms) and the bar angular velocity (pattern speed) in a single snapshot of a galaxy we will look for the stars around the unstable equilibrium points at the bar ends, which move at the same angular velocity than the bar but with less fluctuations than that \citep[see e.g.][]{Athan2003, ChibaFS2021}. Identifying galactic arms without relying on assumptions such as position or density is expected to provide new insights into the formation of spiral arms. We test our technique at times when the bar is fully formed, using two types of simulation data: a test particle simulation where all model components are known, and an N-body simulation, where fewer galactic features are known, but some can be established using multiple snapshots.

In both cases, when applied to snapshots at discrete times, the framework reveals the oscillatory nature of the bar, with stars temporarily belonging to it and changing roles between snapshots. However, it consistently identifies the key components of the galaxy, which are then used to calibrate the potentials of a classical dynamical system model. The resultant invariant manifolds of this model, emanating from the Lyapunov periodic orbits around the unstable equilibrium points, are compared to the arms observed in the simulations.

The paper is organised as follows: in Section~\ref{sec:method} we present the employed techniques from Computational Geometry and Topology discussing their particular application to the identification of galactic features, namely size and pattern speed of the bar, the location of the arms, and the equilibrium points of the system, from a single snapshot. In Section~\ref{sec:manifolds} we demonstrate how to convert the identified features into a dynamic model of the galaxy and illustrate how the equilibrium points of this model define invariant manifolds, which are expected to form the backbone of the spiral arms. In Section~\ref{sec:test} we apply these techniques to two different simulations: a particle test simulation created with a potential from a galaxy with prescribed components, including bar and other features, and an N-body simulation, in which the bar emerges gradually. In both simulations we focus on times where the bar is fully formed. Section~\ref{sec:disc} is devoted to discussing the validity of the obtained results and potential uses of the detection procedure.

\section{Detection of the main galactic features}
\label{sec:method}

Our starting point is a table of positions and velocities in the galactic plane of stars forming the galaxy, at a particular epoch. In disc-shaped galaxies, the galactic plane can be estimated from star positions and velocities in the sky, i.e. without the line-of-sight component, provided that the line-of-sight direction does not meet the galactic plane with a small angle. The population of stars can be the entire (observed, simulated) galaxy, or it can constitute a statistically significative random subsample.

The initial hypothesis is that the galaxy has as distinct major components a bar and a disc, and it may have related features such as arms. We will endeavor to identify these elements, and compute basic properties such as size and pattern speeds. Comparisons, as in Section~\ref{sec:test}, of the observed (or simulated) dynamics with those of the models defined by the detected features, provide evidence for the correctness of the initial hypothesis.

Our techniques purely rely on the analysis of positions and velocities, which turns out to give enough information to determine bar size, pattern speed, position of arms and critical points. Knowledge of the mass of the stars is required for determining densities and centre of mass of the galaxy, or of its components. For those, we assume equal mass for all studied stars, which delivers correct results provided that mass distribution is statistically homogeneous among stars.

\begin{figure}
\begin{center}
\includegraphics[width=1\textwidth]{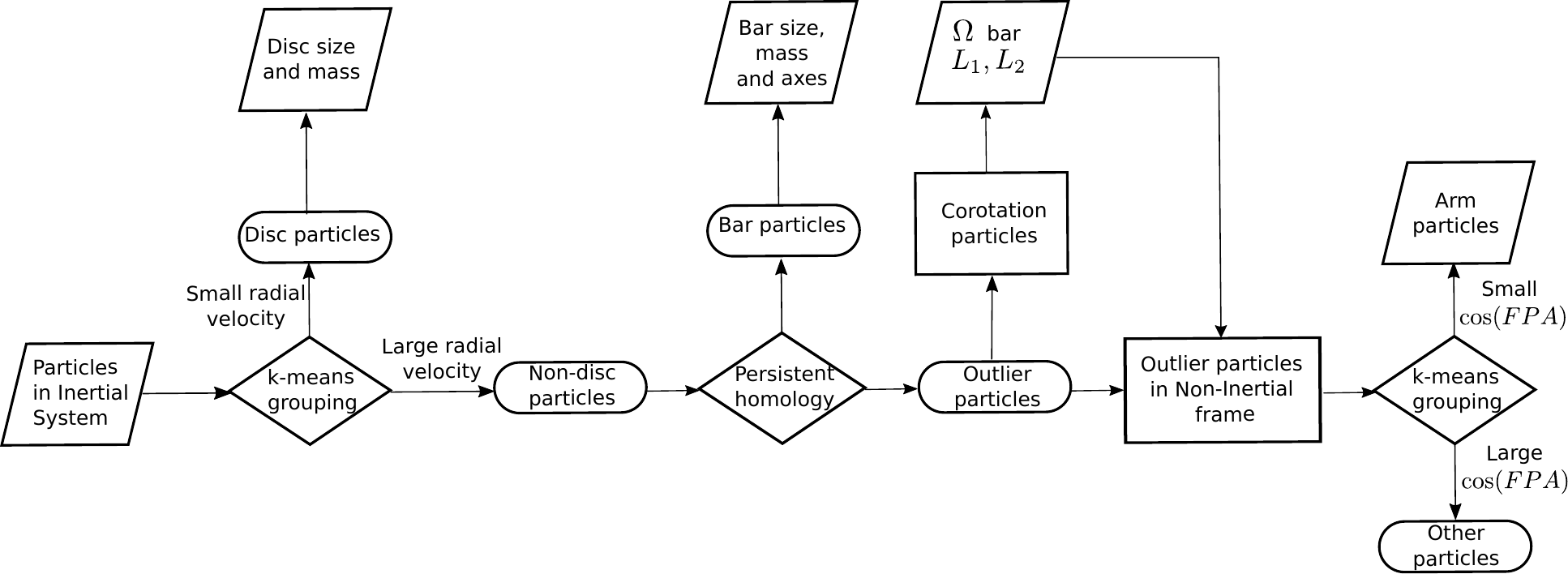}
\end{center}
\caption{Flowchart for capture of short-term galactic characteristics from a single snapshot.}
\label{fig:flowchart}
\end{figure}

For the sake of clarity, the flow chart of the procedure is shown in Fig.~\ref{fig:flowchart}. We will now proceed to detail the proposed method.

\subsection{Identifying temporary bar particles: Radial velocity and topology}

In our procedure, inspired by the $x1$ family of periodic orbits that constitute the backbone of the bar~\citep{Athan1983, Pfenn, Skokos} we first consider a combinatorial analysis of velocities to distinguish between stars we associate to the galactic disc, 
with conditions for a regular rotating motion,
and stars with a more complex motion. Then, we apply a persistent homology analysis in the positions of the second class to separate it in two groups: Stars temporarily belonging to the bar, in the central part of the galaxy, and outliers spread out through the rest of the galaxy.

The information provided by the latest observational projects, such as the data from Gaia, often includes the 3 components of the star velocities. But the line-of-sight component of velocity has usually been estimated with a larger margin of error than the sky ones. Because of this, and the largely flat nature of barred galaxies, such as the Milky Way or the Large Magellanic Cloud, we have found that more accurate results are obtained just using the star velocity components inside the galactic plane, which can be deduced from velocities in the sky, e.g. using the eccentricity of the observed galactic disc. This procedure gives a table of 2--dimensional positions and velocities in the plane, which provides all the necessary information to ascertain bar and disc features.

\begin{figure}
\begin{center}
\includegraphics[width=0.7\textwidth]{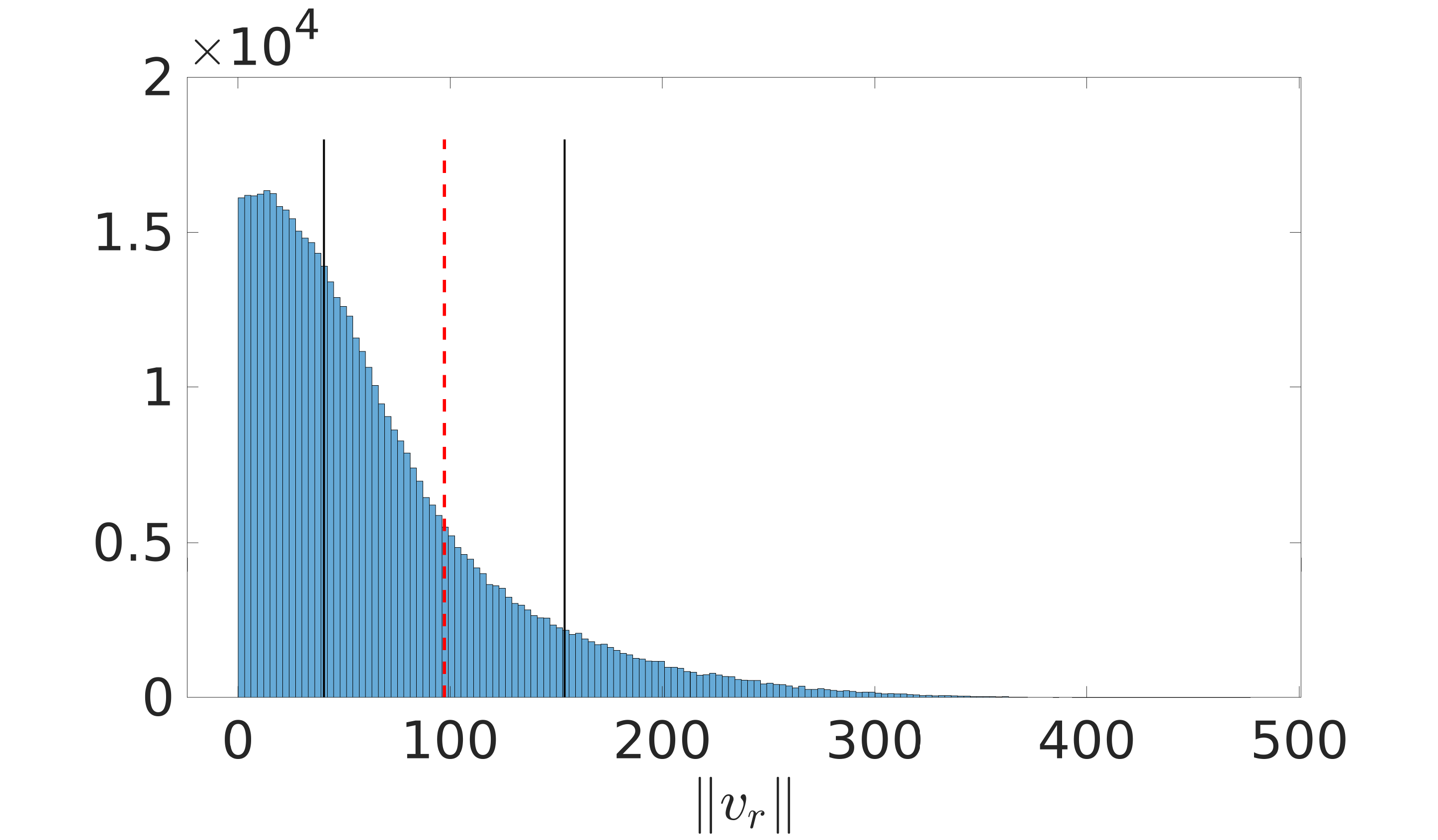}
\end{center}
\caption{Histogram of radial speeds in a star population, subdivided using 2-means grouping (red: partition between the 2
groups; black: average radial velocity for each group).} 
\label{fig:histkmeans}
\end{figure}

Bars that are stable in the long term form primarily in cold discs \citep{DNGFZM2017}. The stars in these discs exhibit simple planar dynamics, rotating around the galaxy's centre in approximately circular orbits. This is the main criterion we use to distinguish them from temporary stars in the bar and arms. In a hotter galactic disc that can still coexist with a bar, disc orbits experience a radial perturbation added to the basic underlying mean circular motion. This radial component introduces noise into our analysis (see Figure \ref{fig:histkmeans}). However, the formation of the bar sets an upper threshold, allowing for a certain margin of noise that is tolerable in our computations.

The initial steps in our procedure are:
\begin{enumerate}
\item Find the centre of the galaxy as the centre of mass of all the stars.
\item For each star, decompose its planar velocity into its radial ($v_r$), and rotational ($v_{rot}$) components, both with respect to the galaxy centre.
\end{enumerate}


Circular motion around the galaxy centre has no radial velocity. Elliptic motion only results in a sizeable radial velocity in cases of large eccentricity. Thus, the stars can be now classified according to their radial speed into 2 groups using the $k$-means grouping algorithm, 
which is a standard algorithm in Computational Geometry (see \cite{DHS2001} for a complete, illustrated description). 

Let us consider a population where each member is characterised by $d$ numeric features, forming a point cloud ${\mathbf X} \subset {\mathbb R}^d$.
This point cloud consists of feature vectors ${X}_m$ for each member $m$. The $k$-means grouping algorithm partitions this point cloud into $k$ groups. The algorithm aims to minimize the sum of distances from each feature vector ${X}_m$ in the cloud to the averaged feature vector of its respective group.
Specifically, for each group $G_i$, this average is calculated as $(\sum_{\tilde m \in G_i}{X}_{\tilde m}) / \#G_i$ where $\#G_i$ denotes the number of elements in group $G_i$.

As illustrated by the example in Fig.~\ref{fig:histkmeans}, we apply this algorithm looking at a single numeric feature: the radial speed.
Stars with smaller radial speed, which is close to 0 on average, follow a rotational (or almost rotational) motion and we associate them to the disc, while the group with larger radial speed is formed by stars with more complex dynamics. 

The stars associated to the bar are part of the group with larger radial speed. By employing topological analysis, specifically using persistent homology on the positions of these stars, we can differentiate this group into two subgroups: stars temporarily in the bar and radial velocity outliers located in other regions of the galaxy.

Persistent homology is also a standard technique in Topological Data Analysis (see \cite{EH2022} for a complete presentation), albeit we are going to consider it in a novel, less standard, way. The standard persistent homology technique consists in the following. Given a point cloud ${\mathbf Q} \subset {\mathbb R}^n$ formed by an arbitrary number of points ${Q}_i$, we consider a {\em maximal distance parameter} $d_{max}$, which is going to move in the range $(0,\infty)$. For each value of $d_{max}$, we look for subsets of points $\{{Q}_0, \dots , {Q}_s\} \subset \mathbf Q$  such that the distance between any two points in the subset is smaller than $d_{max}$. With them, we define a $s$-dimensional simplex which can be thought of as an abstraction of the convex hull of the points in the subset. 

\begin{figure}
\begin{center}
\includegraphics[width=0.7\textwidth]{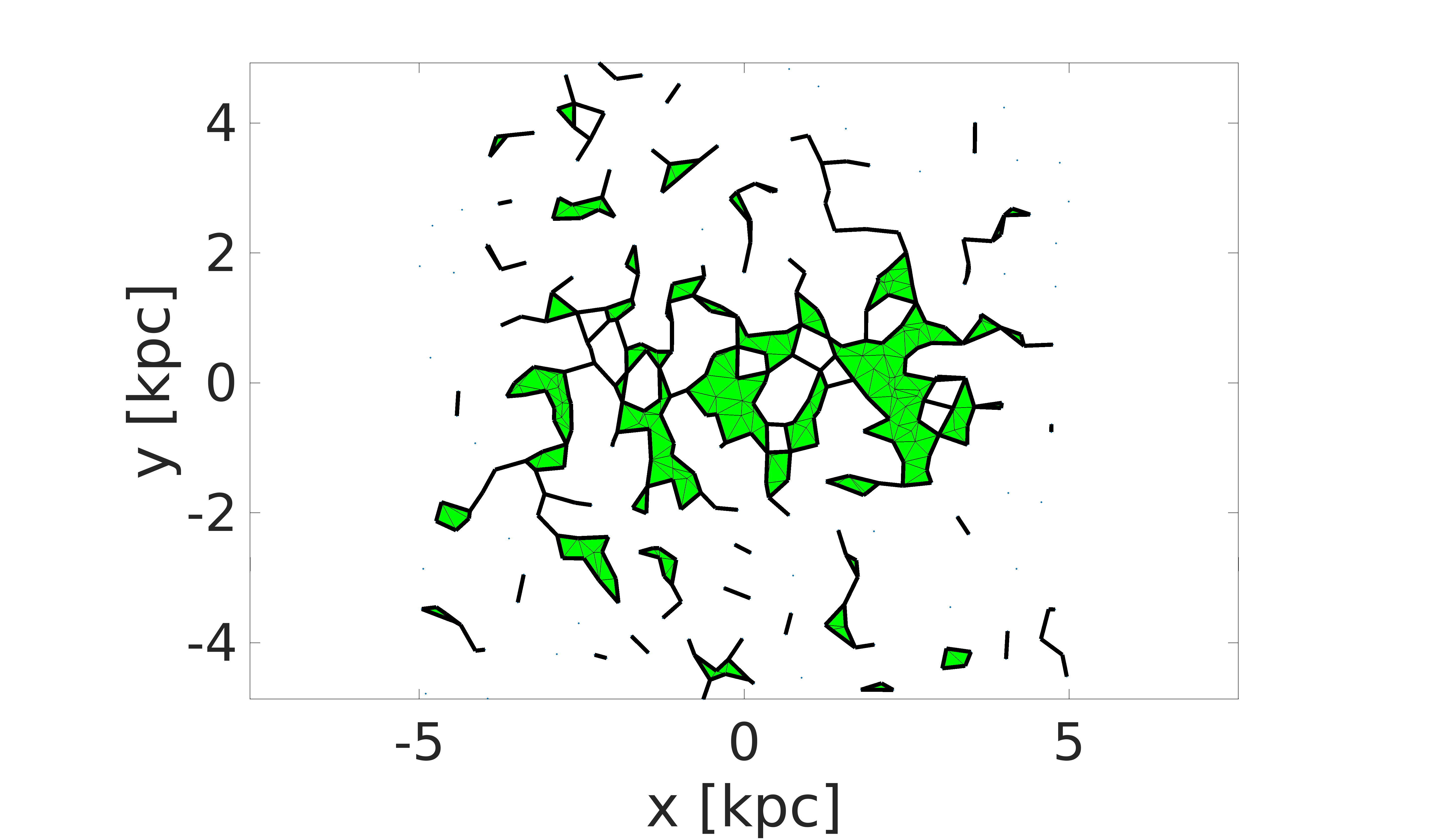}
\end{center}
\caption{Alpha-complex version of persistent homology, with edges and triangles, defined on a random sample of 1000 stars from one of our particle tests, with $d_{max}=0.25$. Note how stars in the denser central region form the largest connected component (i.e. group of stars
connected by edge paths), while stars in the less dense periphery are connected to fewer of their peers by edges.} 
\label{fig:alfa}
\end{figure}

A natural way to reduce both the computational cost and the required background on simplicial polyhedra, is using the Alpha-complex version of persistent homology. In this version, we moreover require that each pair of points in the subset $\{{Q}_0, \dots , {Q}_s\}$ must have neighbouring cells in the Vorono\"{\i} decomposition of ${\mathbb R}^n$ defined by the point cloud ${\mathbf Q}$. Consequences of this additional requirement include that all simplices will have dimension $s \le n$, and that the simplicial polyhedron is the union of the convex hulls of the subsets $\{{Q}_0, \dots , {Q}_s\}$, as illustrated for $n=2$ in Fig.~\ref{fig:alfa}.

Ultimately, persistent homology consists in the analysis of the topology of the polyhedra formed by the point cloud ${\mathbf Q} \subset {\mathbb R}^n$ as the parameter $d_{max}$ ranges from 0 to $\infty$. 

In our case, the point cloud we consider consists of the positions in the galactic plane of the set of stars belonging to the group with high radial speed in the previous $k$-means grouping classification. We consider the Alpha-complex version of persistent homology for this point cloud and the topological feature we study when increasing $d_{max}$ are the connected components of the polyhedron, which are determined by the edges.
So, it suffices to identify the edges rather than to construct the entire Alpha-complex polyhedron. This is, we can ignore general sets of sufficiently close points $\{{Q}_0, \dots , {Q}_s\}$ and just define a graph, in which we join two points in the cloud ${Q}_i,\, {Q}_j$ when their Vorono\"{\i} cells have a common side and $||{Q}_i - {Q}_j|| < d_{max}$.
Overall, the distinguishing feature among galaxy stars with high radial speed is that those temporary associated to the bar cluster densely in the central region of the galaxy, whereas other radial speed outliers are spread out less densely in the galactic periphery.

As the distance threshold $d_{max}$ grows from 0, the connected component of the graph with the largest number of stars appears in the central region of the galaxy because here is where they are more closely packed. Fig.~\ref{fig:alfa} shows an instance of this phenomenon. The number of stars in this connected component grows strongly with $d_{max}$ while the bar coalesces. Once this connected component includes the bar, further growth of $d_{max}$ just incorporates the less dense outliers. This fact turns out in a decreasing slope in the graph accounting for the growth of the number of stars in the connected component. This change in the slope of the growth of the largest connected component is often observed very neatly in the plot (see Fig.~\ref{fig:dmaxpt}). 

In the cases where the transition is not so neat, one may locate it by performing detection of the sharpest corner in the graph. To define this sharpest corner, we rescale the axes in the plot of the number of stars in the largest connected component over $d_{max}$ so that the two axes have the same length (as in Fig.~\ref{fig:dmaxnb}), and then we look for the residue-minimizing fit of a curve in the shape of the letter $\Gamma$ (i.e., of a set square) over the graph.

Two other features have been tested to detect the threshold value of $d_{max}$ delimiting the bar in a sample: 
\begin{itemize}
\item The length of the main bar axis, whose computation is discussed in Section~\ref{s:feat}, successfully detects the bar with this persistent homology procedure in the particle test simulation, and yields somewhat less clear results in an N-body simulation with more complex dynamics.
\item The number of stars with retrograde motion (meaning in the opposite sense to those delineating the disc motion) contained in the largest connected component of the persistent homology graph has been the most successful detector of bar limits in all tested examples. 
\end{itemize}
In all instances, the limit value for $d_{max}$ defining the bar, can be detected as the sharpest corner in the graph obtained with the procedure described above.


When applied to test particle simulations, stopping the persistent homology graph exactly at the point where the growth of the largest connected component has the strongest change of slope, and declaring as bar members the stars inside it, has been found to leave out a few component stars at the tips (along the main axis) of the bar. 
This is due to a combination of 2 facts: the tips of the bar have a much lower density of stars than its core, and the typical orbits of stars in the bar consist of stars following the bar along its main axis and turning around when the end of the bar is approached \citep{Athan1983, Patsis2019, Moges2024}. The radial velocity of the star is low during this turnaround period, so the fewer stars in the tip region of the bar are less likely to be detected by our $k$-means grouping based on the size of the radial speed. This bias in the algorithm is studied in the test particle simulations of Section~\ref{sec:test}. The radial velocity of the star is also low when it is located at the edge near to the minor axis of the bar, but their omission is negligible. Particles near the minor axis move rapidly, meaning that only a small number of particles are affected. Section~\ref{sec:test} further illustrates how, despite these inherent biases, the method achieves enough accuracy in capturing the overall dynamics.


\subsection{Bar features}\label{s:feat}

Once the stars constituting the bar from the galaxy sample have been identified, we can analyse the features of the detected bar. Unlike previous classification methods \citep{Dehnen2023, Pfenniger2023}),
our classification differentiates the stars into two sets, even though they coexist in the same space region. We have observed that this distinction improves the feature estimation.

The shape of the bar in the projection plane is approximately an ellipse, so computation of the Oriented Bounding Box (OBB) through the inertia moment tensor of the point cloud formed by its stars determines its centre, main axes and dimensions.

The instantaneous pattern speed of the bar can be computed from the inertia moment tensor, as explained in \citet{Pfenniger2023}. This pattern speed is the derivative of the argument of the main axis of the bar with respect to time, which can be computed from the positions and velocities of the stars forming it. 

As discussed in Section~\ref{sec:test}, the bar in a galaxy is a largely fluctuating phenomenon with a non-constant pattern speed \citep{Valenz2003, Wu2018}. Consequently, the instantaneous pattern speed of the bar is less relevant to the dynamics of the galaxy than the average pattern speed over a longer time window. Given that our analysis focuses on a single snapshot of the galaxy, we have identified a dynamic feature that is instantaneous, yet more stable than the instantaneous pattern speed and related to a short-term average pattern speed: the particles in corotation. By studying the local density of stars with high radial velocities, we can approximate the positions of the critical points at the ends of the bar. The averaged angular speed of points in these neighbourhoods provides a short-term version of the pattern speed, which fits closely the long-term averaged pattern speed in galactic simulations (as seen in Tables \ref{taula:pt}, \ref{taula:nb}), and is more useful than the instantaneous pattern speed for predicting the galactic dynamics. The fact that we are determining a small area containing each critical point rather than its accurate position is actually a strength of our method: as shown in \citep{Wu2016}, the position of instantaneous equilibrium points with respect to the bar may oscillate, and working with an approximate rather than exact position accounts for these oscillations in the short term.

Let us first discuss the computation of the density of a point cloud with techniques from Computational Geometry. We have considered two different procedures: 

The procedure we use to compute the density of building blocks, such as the bar or the arms, consists in finding 
for each constituting star its $k$ nearest neighbours using the KNN algorithm. This should be done in position space $\mathbb R^3$, but due to the flatness of the galaxy it can be done in the galactic plane. The typical number of sides of Vorono\"{\i} cells in the plane, and sphere packing theory, both suggest $k=6$ as a suitable number of neighbours in a plane. The density of the point cloud at the location of the star is estimated as the number $k$ of neighbours divided by the area of the ball with radius the midpoint between the distances to the $k$-th and $(k+1)$-th neighbours. The numerator $k$ can be replaced by the mass of the neighbouring stars if it is known. The density obtained by means of this procedure closely reflects the local fluctuations in the distribution of stars in the cloud.

The second procedure is used to estimate the average bar pattern speed over a longer time window. In this case, the computed density function sacrifices local resolution in order to gain smoothness and stability in the results.
By fixing a local radius $r_{loc}>0$, we can count the number $N_i$ of points whose distance to a given point ${Q}_i$ is less than $r_{loc}$. Again, this distance is measured in the position space $\mathbb R^3$, and can be measured in the galactic plane because of the flatness of the galaxy. The density of the point cloud about the point ${Q}_i$ can be estimated as the quotient $N_i/ V_{r_{loc}}$, where $V_{r_{loc}}$ is the measure of the ball of radius $r_{loc}$ in ambient space, namely in the galactic plane. If the mass of the points is known, it can be readily incorporated to the computation. A larger value for the local radius $r_{loc}$ results in a smoother density function, which can actually be more accurate if the masses of the stars are unknown and unequal, but follow a normal, or uniform, random distribution.  

With a view towards the case of real galaxies, where individual star masses are not precisely known, we have computed densities of stars in the galactic plane using local radius ranging from 1/12 to 1/200 of the bar length. So in the areas at the ends of the bar where the critical points $L_1$, $L_2$ are to be expected, the local ball around each star contains $0.03-0.5\%$ of the stars. This procedure can be understood as an average over the actual density of the galaxy. 

Following the second procedure of density determination, we take the stars in the galactic plane which are classified as radial speed outliers after the determination of the bar. For each one of them, we compute two densities: 
\begin{enumerate}
\item The density $\rho_{out}$ of the point cloud formed exclusively by outlier stars,
\item The density $\rho$ of the entire point cloud.
\end{enumerate}
These two densities are means to find the relative density $\rho_{rel}=\frac{\rho_{out}}{\rho}$, i.e. the fraction of neighbours at each radial speed outlier point which are also radial speed outliers. 

There is a direct relation between the position of the critical points $L_1$, $L_2$ at the ends of the bar in a barred galaxy and the relative density of radial speed outliers we have just introduced. In a rotating reference frame attached to the bar, the force field vanishes at the critical points, and has the smallest values of the modulus in their neighbourhood. This means that the neighbourhood of the critical points $L_1$, $L_2$ holds particles moving slowly (in the non-inertial reference frame), and consequently remaining longer there than in regions further from them. Static (resp. slow moving) particles in the non-inertial frame mean particles with a circular (resp. almost circular) motion in the inertial reference frame. Therefore, the critical points $L_1$, $L_2$ at the end of the bar are expected to lie around a local minimum of the relative density $\rho_{rel}$ of radial speed outliers. These local minima of $\rho_{rel}$ ought to have a large attraction basin, and rotate with the bar, i.e. with a comparable pattern speed. In other words, this means we seek particles in corotation with the bar, as these particles provide a more stable, long-term, pattern speed estimation of the bar.

The above analysis is the basis of the procedure we follow to determine a long-term version of the bar pattern speed which reflects its dynamics in a stable way: 
\begin{enumerate}
\item Select a square box at each end of the bar, aligned with the bar axis and with edge 1.5 times the bar semiaxis,
\item Look for the minimum of the relative density $\rho_{rel}$ in this box; this is the approximate position of an equilibrium point,
\item Take the outlier stars with non-retrograde motion at a distance under 1/48 of bar length of this minimum, and average their angular velocities in the inertial reference frame.
\end{enumerate}
The average of angular velocities over the outliers in the neighbourhoods of the points $L_1$, $L_2$ provides an estimate of the pattern speed of the bar which better reflects the dynamics of the galaxy in the studied particle tests and $N$-body simulations.

This procedure depends on three parameters: The density computation radius $r_{loc}$, the edge size of the box containing the critical points, and the radius for averaging angular velocities around each critical point $L_1$, $L_2$. The location of the critical points $L_1$, $L_2$ has been found to be stable with respect to variations in $r_{loc}$ within the above specified range of values, as well as with variations in box size, provided the box is large enough to contain the critical points. This can be easily inferred, for example, if there are galactic arms starting at these critical points. The radius for averaging of angular velocities has been selected by starting from a very small radius, such as the distance from the selected critical point to the sixth neighbour, and increasing the averaging radius until stable results are achieved.


\subsection{Arm detection}

A barred galaxy may have arms originating from the tips of its bar, transporting matter from its origin to the galactic periphery. The detection of particles belonging to the arms relies on the fact that, since their trajectory follows the arm, their motion is neither circular nor elliptical with low eccentricity, as it is in the galactic disc.
This motion favours the hypothesis that the existence of arms can be explained through the Invariant Manifold Theory \citep[see e.g.][]{Romero1, Romero2, Warps, Asymmetry}. In this theory, the two equilibrium points located at the ends of the bar have associated Lyapunov periodic orbits from which invariant manifolds emanate acting like backbone structures. 
These structures transport matter from the neighbourhood of each equilibrium point to the periphery of the galaxy like channels, with the added stretching property of the unstable manifolds \citep{LCS2018} that make them visible in the form of spiral arms, or in the form of a ring galaxy, when they approach to the opposite equilibrium point.
 
In a first step, let us change positions and velocities from the original inertial reference to the synodical (non-inertial, rotating) reference frame, where the $x$ axis is aligned with the major axis of the bar. This reference preserves the origin but rotates its axes with the pattern speed of the bar, so that the position of the bar becomes static in time \citep{Romero1, Asymmetry}.

Then, for each star we compute the Flight Path Angle (FPA, defined as the angle measured in the synodical frame between the velocity of the star and the tangent to a circular orbit around the galaxy centre passing through the star). We use again the $k$-means grouping algorithm to partition the set of stars outside the region of the bar in 2 groups using the $|\cos(FPA)|$ feature. The group of stars with a high value of $|\cos(FPA)|$ are following approximately circular orbits, with the sign of $\cos(FPA)$ largely irrelevant because it reflects whether their angular speed is greater or smaller than the computed pattern speed for the bar.
On the other hand, the stars with a low value of $|\cos(FPA)|$ tend to follow noncircular orbits and are labelled as members of the galactic arms.

This detection scheme for galactic arms is less precise than the one proposed for the bar, as it tends to miss stars in parts of the trajectory of the arms that are orthogonal to the radius. Additionally, the simulations used to test our method assume equal particle masses. Under these conditions, the method performs as described in the results. If the particle masses were known, we could compute their total energy, leading to a more accurate identification of the equilibrium points of the system. This, in turn, would improve the precision of our detection of galactic arms, which emanate from around the unstable equilibrium points and transport matter within a specific range of energy levels \citep{Romero1, Warps, Asymmetry}.

Despite of the significant noise introduced by the lack of knowledge about the star masses, a statistical study shows that the proposed arm detection scheme provides valuable information about the existence and position of galactic arms. To reduce the fluctuations caused by varying energy levels, we replace the velocity in the non-inertial reference frame of each star by the average velocity of nearby stars, where ``nearby'' refers to stars within a fixed distance $d$. By using velocities averaged over a distance $d$ equal to $\frac{1}{16}$ of the length of the bar, the 2 pairs of arms connecting the critical points at the tips of the bar, transporting matter in opposite directions, become clearly visible in the plot of the velocities of arm particles in the non-inertial frame (see Fig.~\ref{fig:test_barrabras}).

As an additional verification of our arm detection scheme, we can re-detect the equilibrium points $L_1$, $L_2$ at the tips of the bar, as well as the equilibrium points $L_4$, $L_5$ in the zero velocity regions above and below the bar, surrounded by the arms in the plane.
These points of the dynamical system are characterised by the property that, at the energy level of the equilibrium point, the velocity vanishes in the non-inertial frame.

Applying the same ideas used in the determination of the pattern speed of the bar, now we can take the stars detected as belonging to the arms, along with their averaged velocities, and search for the star with minimum speed in each of the four plane sectors delimited by the diagonals defined by the axes of the bar.
The locations of these stars track the positions of the equilibrium points of the system. 

As we show in the results of Section~\ref{sec:test}, the positions of $L_1$, $L_2$ found using this method match, both in value and variance, those we find when determining the pattern speed of the bar. However, the positions found for the points $L_4$, $L_5$ exhibit high variance.
The points $L_4$, $L_5$ lie within the Hill regions for several energy levels, resulting in slow motion and low density of stars within these regions.

.\section{Dynamical Model}
\label{sec:manifolds}

The methodology described in Sect.~\ref{sec:method} provides values that can be used to calibrate a galaxy model and generate simulations within the mathematical model afterwards. For this purpose, we consider a classical dynamical system that describes the motion of a particle under a gravitational potential $\phi$ and a rotating frame aligned with the main axis of the bar \citep[see e.g.][]{Pfenn, Skokos, Romero1, Warps, Asymmetry}:

\begin{equation}
\left\lbrace
\begin{array}{l}
 \ddot{x} = \phantom{-}2\,\Omega\, \dot{y} + \Omega^2\, x  - \phi_{x} \\
 \ddot{y} = -2\,\Omega\, \dot{x} + \Omega^2\, y - \phi_{y} \\
 \ddot{z} = - \phi_{z}\,.\\
\end{array}
\right.
\label{eqn:systmodel}
\end{equation}

The total gravitational potential $\phi$ of the system is created by the different components of the galaxy (named bar, disc and in the test particle simulation, also halo). The size, mass and parameters of each component is going to be explicitly detailed in the corresponding simulation. 

We model the potential of the bar, $\phi_b$, by means of a Ferrers ellipsoid \citep{Ferrers} with density function,
\begin{equation}\label{eqn:Ferrers}
 \rho = 
 \left\lbrace
 \begin{array}{ll}
  \rho_0(1-m^2)^{n_h}, & m\leq 1, \\
  0,   & m > 1, \\
 \end{array}
 \right.
\end{equation}
where $m^2=x^2/a^2 + y^2/b^2 + z^2/c^2$, and \emph{a} (semi-major axis), \emph{b} (intermediate axis) and \emph{c}  (semi-minor axis) determine the shape of the bar. The value $n_h$ is the homogeneity degree of
the mass distribution ($n_h=1$ in our work) and $\rho_0$ is the density at the origin ($\rho_0=\frac{105}{16\pi}\frac{GM_b}{abc}$ if $n_h=1$, where $M_b$ is the mass of the bar).

We describe the disc potential, $\phi_d$, by means of the equation \citep{Miyamoto}
\begin{equation}\label{eqn:Miyamoto}
 \phi_d = - \frac{GM_d}{\sqrt{R^2+(A+\sqrt{B^2+z^2})^2}},
\end{equation}
were $R^2=x^2+y^2$ is the cylindrical coordinate radius of the potential in the disc plane and $z$ is the vertical distance over the disc component. The parameters $G$, $M_d$, $A$ and $B$ denote the gravitational constant, the disc mass and the shape of the disc, respectively. 

Occasionally, we consider a halo to describe spherical dark matter with potential \citep{Allen1991} 
\begin{equation}\label{eqn:halo}
\begin{split}
 \phi_h = &-\left[\dfrac{GM_h(R/a_h)^{2.02}}{R\left(1+(R/a_h)^{1.02}\right)}\right] \\
 &-\dfrac{GM_h}{1.02a_h}\left[-\dfrac{1.02}{1+(R/a_h)^{1.02}}+\ln\left(1+\frac{R}{a_h}\right)^{1.02}\right],
 \end{split}
\end{equation}
where the parameters $M_h$, and $R$ denote the halo mass and its radius, respectively. The parameter $a_h$ is set to fit the rotation curve of the galaxy.

The unit of length we consider is the kpc, the time unit is u$_t = 2 \times 10^6$~yr,  $\Omega$ is in~[u$_t$]$^{-1}$, and the mass unit is u$_\text{\tiny{M}} = 2 \times 10^{11} M_\odot$, where $M_\odot$ denotes the mass of the Sun. We take the convention that $G(M_d+M_b+M_h) = 1$. Note that the bar pattern speed unit in the simulations is km s$^{-1}$ kpc$^{-1}$, and $\Omega=0.05$~[u$_t$]$^{-1} \approx 24.46$~km s$^{-1}$ kpc$^{-1}$.

As usual, the effective potential of the model is given by $\phi_{_{\hbox{\scriptsize eff}}} = \phi - \frac{1}{2}\,\Omega^2\, (x^2 + y^2)$ and we define the Jacobi first integral as, 
 \begin{equation}\label{eqn:PreCJAC}
   C_J(x,y,z,\dot{x},\dot{y},\dot{z}) = -\,(\dot{x}^2+\dot{y}^2+\dot{z}^2)+\Omega^2\,(x^2+y^2)-2\,\phi,
\end{equation}
which can be thought as the energy of a trajectory in the rotating frame.

The analytical model~\eqref{eqn:systmodel} has five equilibrium points located at the positions which fulfill $\nabla\phi_{_{\hbox{\scriptsize eff}}} = 0$. They are usually called Lagrangian points and denoted by $L_i$, $i=1\ldots 5$. The collinear equilibrium points $L_1$ and $L_2$ are linearly unstable and they are located at the ends of the bar, aligned with its main axis, while $L_4$ and $L_5$ are linearly stable and placed also in the plane $z=0$ but outside of the $x$-axis. Additionally, $L_3$ appears on the origin of coordinates and it is also linearly stable.  

From the unstable equilibria $L_1$ and $L_2$ there emanate two families of Lyapunov periodic orbits, the planar and the vertical families, both unstable in a neighbourhood of these points. Vertical Lyapunov orbits are not relevant for the transport of matter between the regions delimited by the zero velocity curves because they mostly add a vertical oscillation \citep{Rom09}. From each planar Lyapunov orbit there emanate stable and unstable invariant manifolds. The stable manifold consists of the set of orbits that asymptotically tend to the periodic orbit forward in time while the unstable manifold is determined by the orbits that depart asymptotically from the periodic orbit. Visible arms and rings structures in galaxies are driven by the set of these unstable manifolds.
More detailed explanations about the dynamics around these points can be found in \citet{Athan1983, Romero1, Warps, Asymmetry}.

Moreover, the zero velocity surfaces $\phi_{_{\hbox{\scriptsize eff}}}=C_J$ bound the region of possible motion for a given energy. In particular, the $(x,y,z)$ regions where $\phi_{_{\hbox{\scriptsize eff}}}(x,y,z)>C_J$ are known as Hill regions, which are forbidden regions for a star with energy $C_J$.

\section{Testing cases}
\label{sec:test}

In this section, we validate the proposed methodology for detecting the key characteristics of barred galaxies using two distinct types of simulations: a test particle simulation and an N-body simulation. In both cases, the results are consistent. In the test particle simulation, all characteristics of the model are known, while in the N-body simulation, fewer galactic features are initially available, but some other can be identified by analysing multiple snapshots at successive time intervals. This comparison allows us to evaluate how the results obtained from a single snapshot at a particular time, align with the galaxy evolutionary behaviour, depicted by previous and latter snapshots.

\subsection{Model calibration from a test particle simulation} \label{ss:pt}

The test particle simulation has been generated using the same method as in \citet{Rom15}. The potential in this simulation is created by:
\begin{itemize}
\item A Ferrers bar, Eq.~\eqref{eqn:Ferrers}, with potential $\phi_b$, and with semi-major axis $a=4.5$ kpc, intermediate axis $b = 0.675$ kpc, semi-minor axis $c = 0.117$ kpc and mass $M_b = 2.5 \times 10^9$ $M_\odot$.
\item A short bar or CORBE/DIRBE bulge, with potential $\phi_{sb}$ modeled as a Ferrers ellipsoid, Eq.~\eqref{eqn:Ferrers}, with a semi-major axis $a = 3.13$ kpc, axis ratios to $b/a = 0.4$ and $c/a = 0.29$, and mass $M_{sb} = 4.5 \times 10^9$ $M_\odot$.
\item A Miyamoto-Nagai disc, Eq.~\eqref{eqn:Miyamoto}, with potential $\phi_d$, and with parameters $A = 5.3178$ kpc, $B = 0.25$ kpc and $M_d = 8,5 \times 10^{10}$ $M_\odot$. 
\item A dark matter halo, Eq.~\eqref{eqn:halo}, with spherical potential $\phi_h$, extending to a maximum radius $R_{max} = 100$ kpc and $M_h = 10.7 \times 10^{10}$ $M_\odot$. 
\end{itemize}
The values of the parameters for the disc, halo and bulge have been taken from \citet{Allen1991}. The total potential of the model is the sum of these building block components, $\phi = \phi_d + \phi_b + \phi_{sb}+\phi_h$.

The pattern speed of the bar is $\Omega = -25 $ km s$^{-1}$ kpc$^{-1}$, corresponding to a rotation period $T=78.2 \, \pi$ Myr $\simeq 245.8$ Myr. The equilibrium points of the system have approximate locations $L_1\approx(4.5,0)$, $L_2\approx(-4.5,0)$, $L_3\approx(0,0)$, $L_4\approx(0,4)$, $L_5\approx(0,-4)$.

The simulation uses 10 million particles, gradually introduced to form the bar adiabatically. Snapshots are available at times $t=0, 2T, 4T,16T, 32T$. The galaxy components are well-formed in the last 2 snapshots, and fairly-formed in the previous one. 

The same analysis has been performed for each of the 2 last snapshots: 100 batches of $5 \times 10^5$ particles each have been selected randomly. Batch sizes of $2.5 \times 10^5$ and higher have been found sufficient enough for feature detection. 

\begin{figure}
\begin{center}
\includegraphics[width=0.32\textwidth]{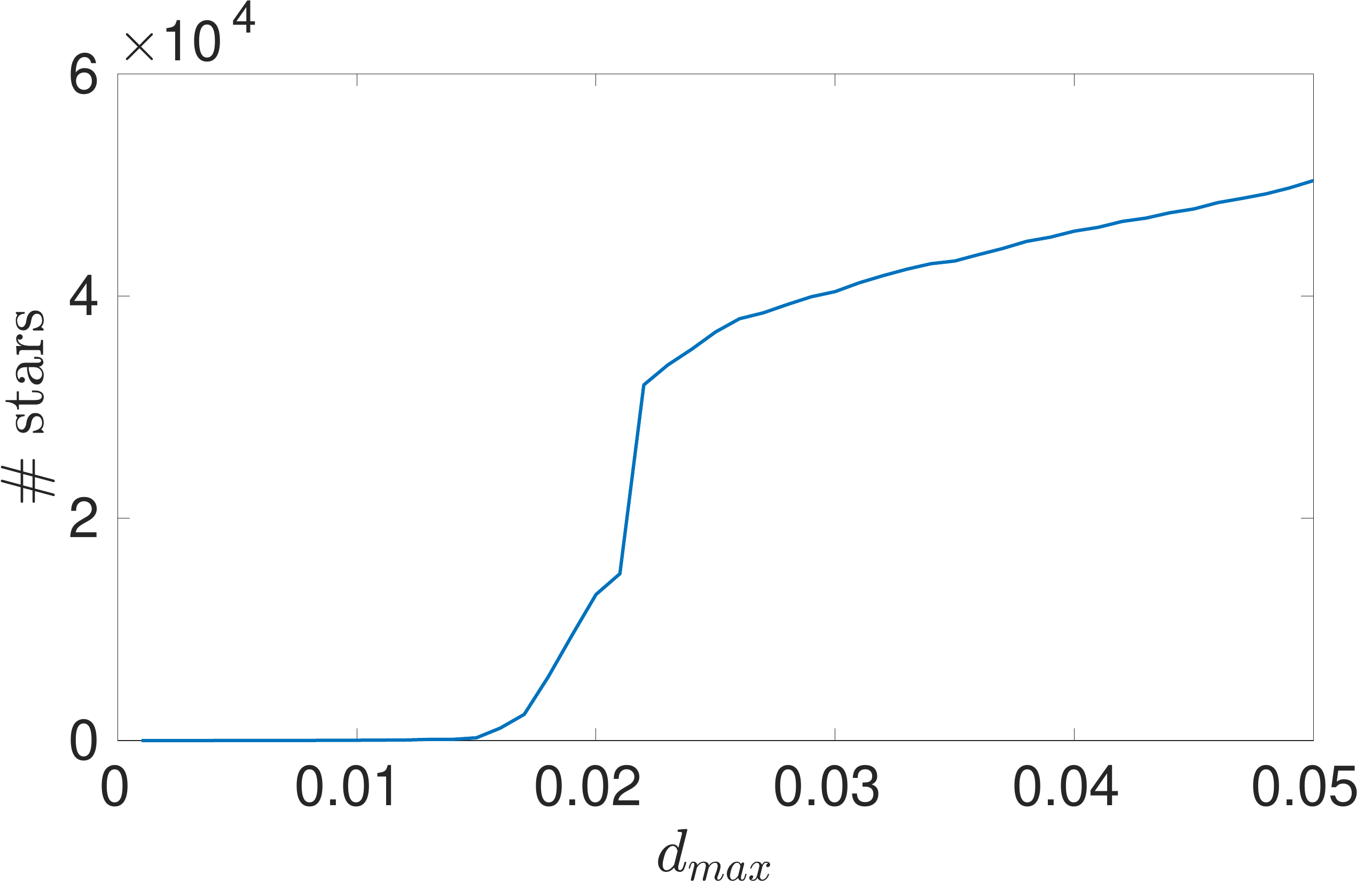}
\includegraphics[width=0.33\textwidth]{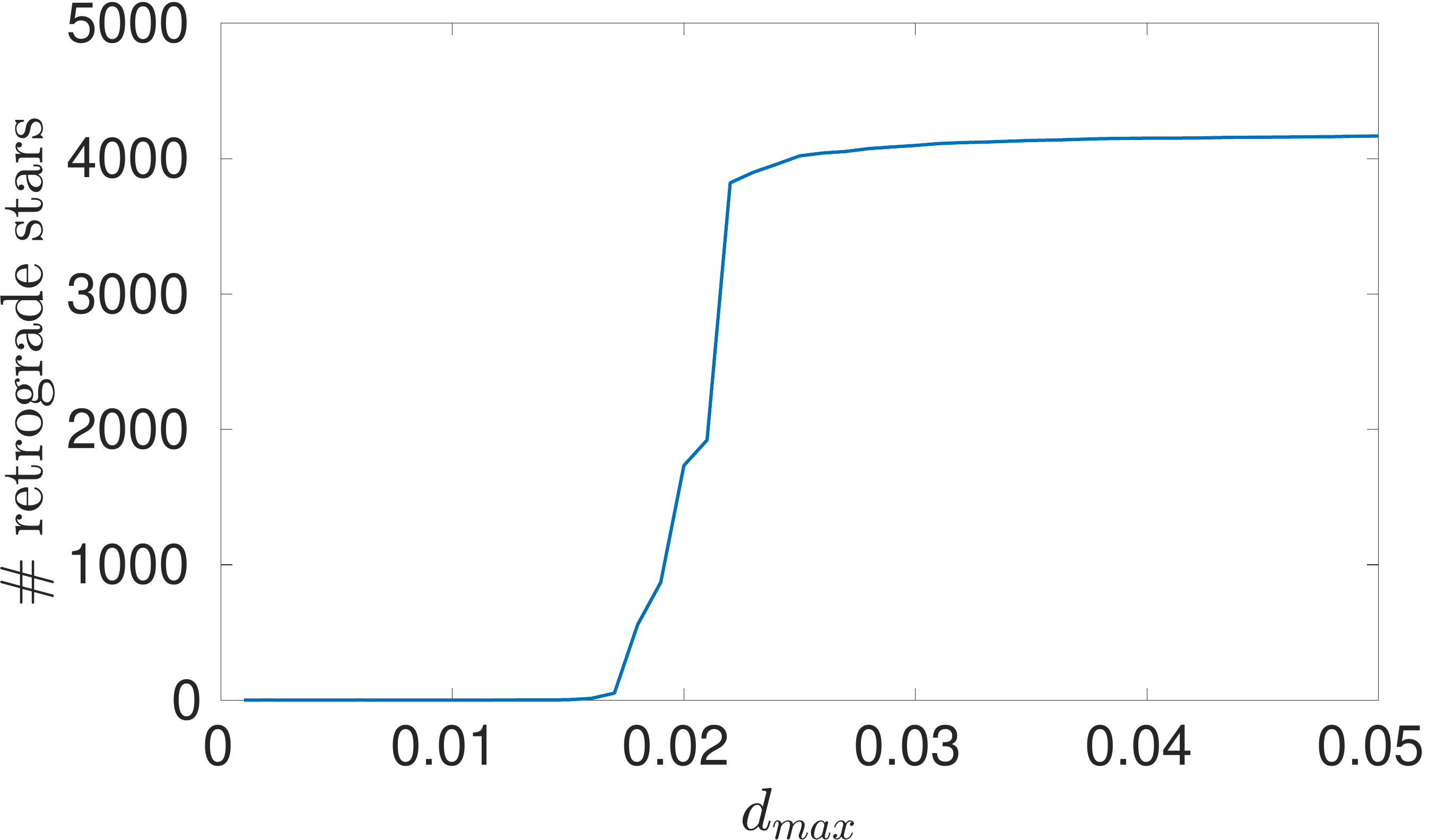}
\includegraphics[width=0.32\textwidth]{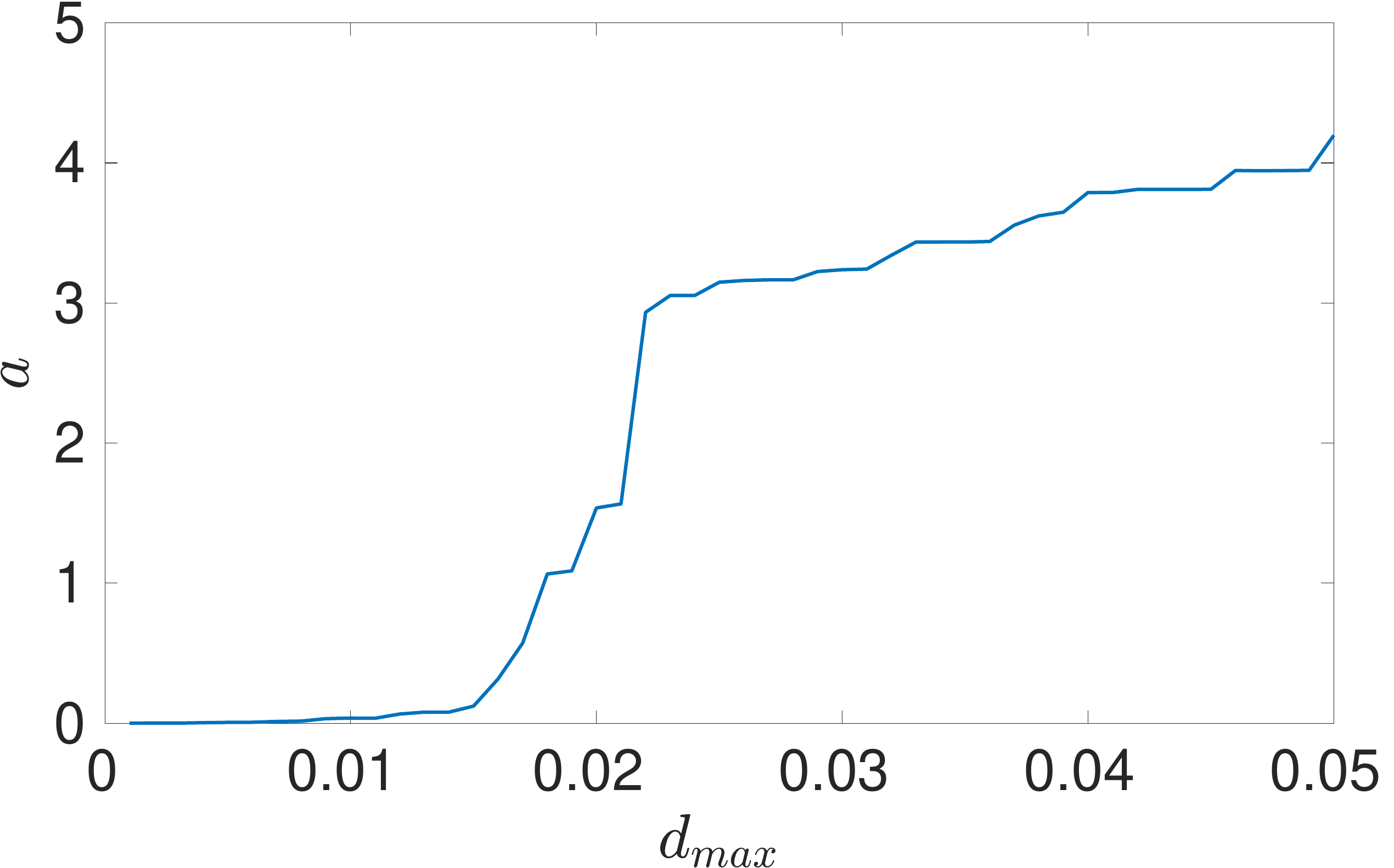} 
\end{center}
\caption{Sample batch 1 from Particle Test Simulation at time $32T$: growth of the connected component identified as the bar with respect to the persistent homology parameter $d_{max}$: (Left) in number of stars; (centre) in number of stars with a retrograde motion; (right) in main semiaxis length $a$. In the three cases, the change in slope best fitting the shape $\Gamma$ happens at $d_{max}=0.022$.} \label{fig:dmaxpt}
\end{figure}

The detection of the bar has been performed on each batch. The stopping criterion for the persistent homology parameter $d_{max}$ yield consistent results across the 3 criteria described in Section~\ref{sec:method}, as it is shown in Figure~\ref{fig:dmaxpt}, for a typical batch.

After the determination of the bar, we estimate its axes, size and pattern speed as described in Sect.~\ref{sec:method}. The pattern speed we obtain from the simulation (rather than its exact value which is known a priori) is the one we select to define the non-inertial reference frame and to detect the arms and estimate the position of the critical points.

The results we obtain for the bar axis, size, pattern speed, and the position of the critical points outside the bar (which we use as a proxy for our estimation of arm existence and position) considering all the data batches of snapshots, provide a mean value and the standard deviation for each feature. All these magnitudes and coordinates have been found to follow a normal distribution according to the Kolmogorov-Smirnov normality test. 

The results for the last 2 snapshots are summarised in Table~\ref{taula:pt}. The method accurately detects the short bar by applying the strictness criterion described in Section~\ref{s:feat}. This criterion enables the detection of the bar pattern speed with high precision. For the detection of the second (large) bar, we extract the particles labeled as short bar from the simulation data and apply the method again. For simplicity, Table~\ref{taula:pt} does not show the large bar, its detected size is $a=3.7751$~kpc, $b=1.5969$~kpc. The sum of the masses we find for the two bars is $10\%$ (the test particle simulation model was created with the total mass of bars set to $6.8\%$). 

\begin{figure}
\begin{center}
\includegraphics[width=0.5\textwidth]{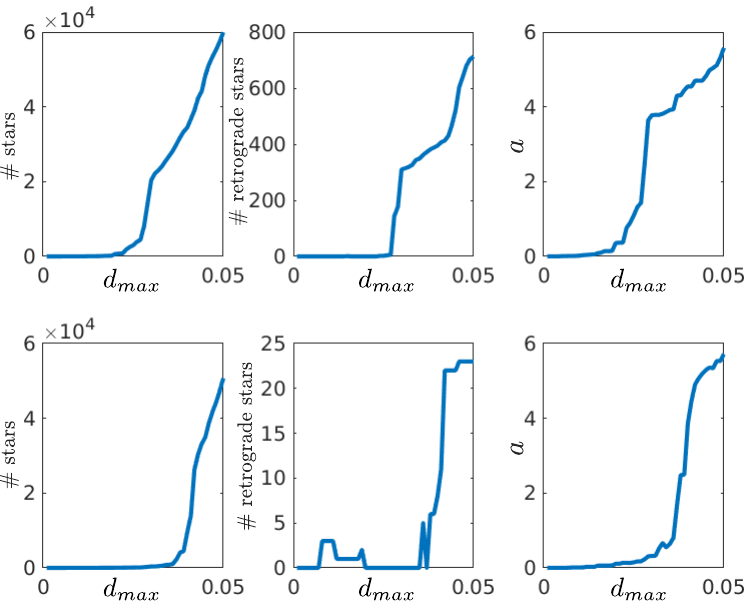}
\end{center}
\caption{Sample batch 1 from Particle Test Simulation at time $16T$: growth of the connected component identified as the bar with respect to the persistent homology parameter $d_{max}$ during the detection of the second bar (top row), and during the search for a third bar (bottom row): in number of stars (left); in number of stars with a retrograde motion (centre); in main semiaxis length $a$ (right). Note the erratic behaviour of the plot of retrograde stars in the second row.
It means that, rather than a single largest connected component being established early and subsequently growing, several connected components of comparable size appear. They compete to be the largest one and do not merge until very late.} \label{fig:2a3abarres}
\end{figure}

While our method detects well a second bar, it also rules out the existence of a third bar in the galaxy: the graphs showing the growth of the largest connected component which constitute the bar, versus the persistent homology parameter $d_{max}$, behave as in Fig.~\ref{fig:dmaxpt} during the detection of the second bar. They show an steady growth and an optimal value for $d_{max}$, defining the bar. On the other hand, the same plots display a more erratic behaviour during the detection of a third bar. Notably, the number of retrograde stars in the largest connected component goes up and down repeatedly, which means that several large connected components are coalescing instead of a single one, as it would be the case if a central bar was being detected. Moreover, these distinct connected components are sufficiently far apart so that they do not merge until $d_{max}$ is very large (see Fig.~\ref{fig:2a3abarres}).

\begin{table*}
\resizebox{\textwidth}{!}{%
        \begin{tabular}{|c|c|c|c|c|c|c|c|}
                \hline
                Time [Gyr] & $a$ [kpc] & $b$ [kpc] & $\Omega$ [km s$^{-1}$ kpc$^{-1}$] & L1 & L2 & L4 & L5  \\ \hline
                $16T$    & 3.0215 $\pm 0.0740$ & 1.0586 $\pm 0.0665$ & -24.0548 $\pm 1.5120$ & 
$\begin{pmatrix} 4.2207 \\ $-0.0272$ \end{pmatrix} \pm \begin{pmatrix} 0.4143 \\ 0.1292 \end{pmatrix}$ &
$\begin{pmatrix} $-4.2265$ \\ $-0.0040$ \end{pmatrix} \pm \begin{pmatrix} 0.4049 \\ 0.1592 \end{pmatrix}$ &
$\begin{pmatrix} $-0.0480$ \\ 4.0628 \end{pmatrix} \pm \begin{pmatrix} 0.2413 \\ 0.2813 \end{pmatrix}$ &
$\begin{pmatrix} 0.0155 \\ $-4.0543$ \end{pmatrix} \pm \begin{pmatrix} 0.2944 \\ 0.2811 \end{pmatrix}$ \\ \hline
                $32T$    & 2.9537 $\pm 0.0750$ & 1.0497 $\pm 0.0596$ & -23.6026 $\pm 1.7631$ & 
$\begin{pmatrix} 4.4140 \\ 0.0259 \end{pmatrix} \pm \begin{pmatrix} 0.4792 \\ 0.1378 \end{pmatrix}$ &
$\begin{pmatrix} $-4.3711$ \\ -0.0072 \end{pmatrix} \pm \begin{pmatrix} 0.4940 \\ 0.1435 \end{pmatrix}$ &
$\begin{pmatrix} $-0.0345$ \\ 4.1506 \end{pmatrix} \pm \begin{pmatrix} 0.2670 \\ 0.3273 \end{pmatrix}$ &
$\begin{pmatrix} 0.0140 \\ $-4.0543$ \end{pmatrix} \pm \begin{pmatrix} 0.2687 \\ 0.3242 \end{pmatrix}$ \\
\hline 
        \end{tabular}}
        \caption{Detected bar features in the Particle Test Simulation, whose exact pattern speed is $-25$~km s$^{-1}$ kpc$^{-1}$. All variables follow normal distribution. For equilibrium points, the standard deviation applies separately to each coordinate.} 
        \label{taula:pt}
\end{table*}

In order to see whether the stars forming the bar and arms are essentially the same at all times or they belong temporarily to the bar, we take the same random selection of stars to define the batches for the 2 snapshots of the test particle simulation at times $t_4=16T$, $t_5=32T$. The results for a typical batch are summarised in Table~\ref{taula:corrpt}, which shows that belonging to either the bar or the arm is a transitory phenomenon.

\begin{table}
        \centering
        \begin{tabular}{|c|c|c|c|}
                \hline
                Snapshots & $t_4=16T$ & $t_5=32T$ & $t_4$ and $t_5$ \\ \hline
                bar    & 29481 & 32011 & 11686  \\ \hline
                arms   & 66149 & 62207 & 19920 \\ \hline
        \end{tabular}
        \caption{Correlation table for belonging to the bar and arms in batch 1 of stars of the particle test. 1st row: Stars belonging to the bar in snapshots of the test particle at times $t_4,t_5$, stars belonging to the bar in both snapshots at times $t_4$ and $t_5$. 2nd row: Same for arms.}
        \label{taula:corrpt}
\end{table}

\begin{figure}
\centering
\includegraphics[width=0.7\textwidth]{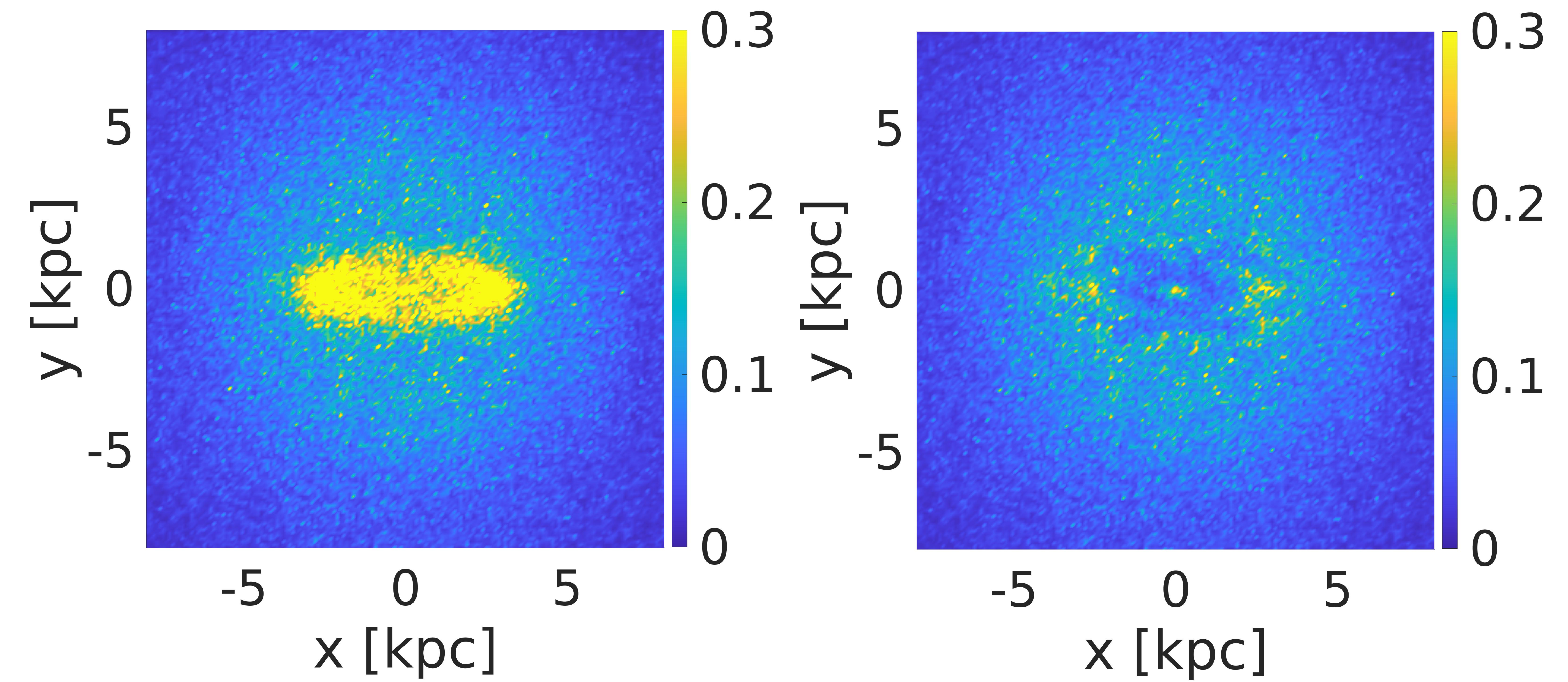}
\caption{Example from the test particle simulation. Left: Density plot of all the particles in the simulation. Right: Density plot of all particles without the bar particles.} \label{fig:test_denstot_y_disco}
\end{figure} 

The subsequent illustrations depict the outcomes of our methodology for identifying the main features of the galaxy as applied to the test particle simulation. Fig.~\ref{fig:test_denstot_y_disco} presents density plots of all particles within the simulation, computed using the 6 nearest neighbours.
The left and right panels exhibit the overall density plot, and the same plot excluding bar particles, respectively. It is notable that a substantial portion of particles remains within the central region once the bar particles are excluded.

\begin{figure}
\centering
\includegraphics[width=0.7\textwidth]{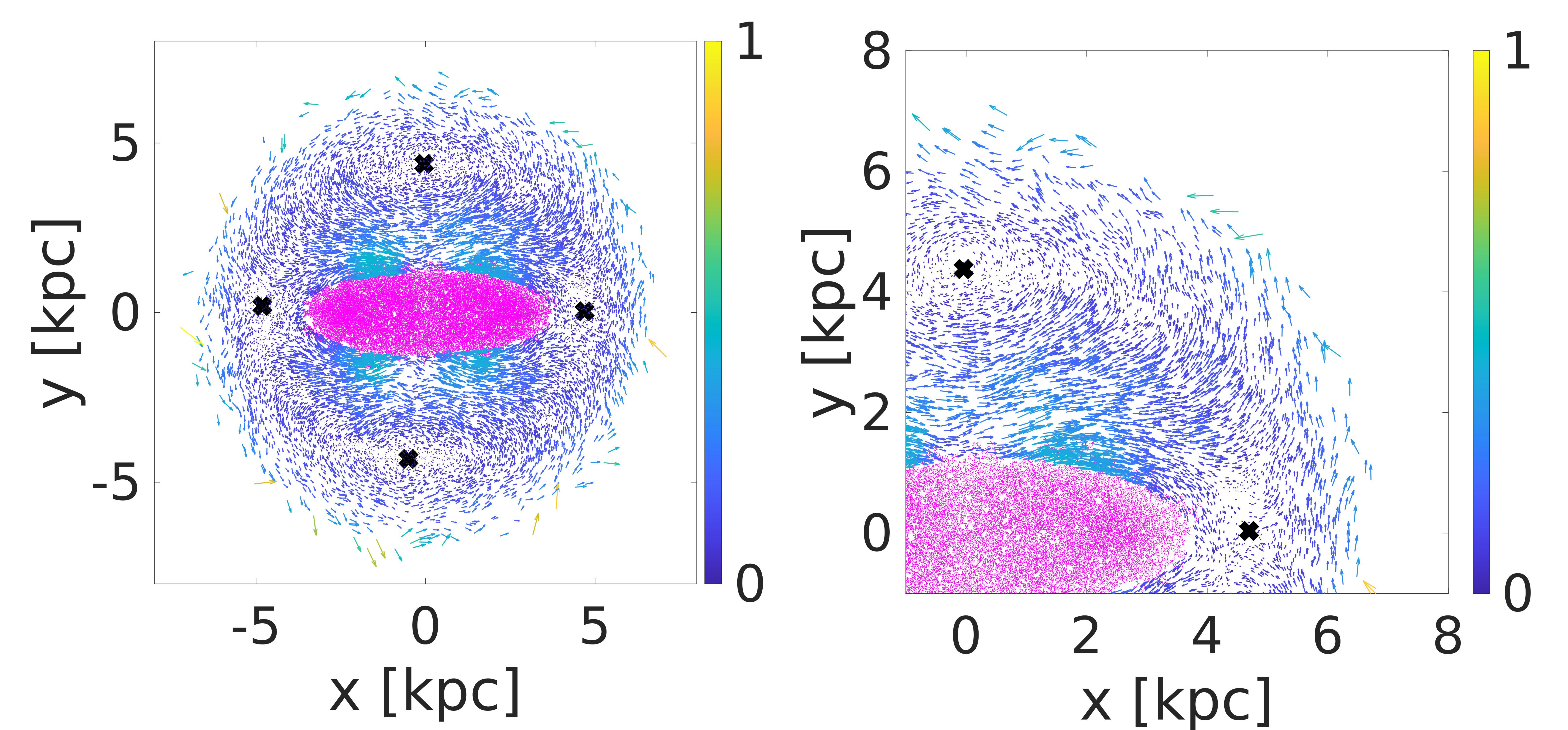}
\caption{Example from the test particle simulation. Left: Stars of the bar (in magenta); averaged velocity vectors of stars forming the arms (with magnitude according to the color bar): note that the stars in the inner arms move in a direction which is the opposite from those in the outer arms; estimated position for the equilibrium points (in black). Right: Zoom of the right quadrant.
}\label{fig:test_barrabras}
\end{figure}

Fig.~\ref{fig:test_barrabras} illustrates the averaged velocity vectors of stars constituting the arms in the non-inertial reference system, along with the estimated position of the equilibrium points (in black). The region characterised by zero velocity curves, where particle movement is constrained for a given energy level, is also distinguished, with vectors displaying a near zero modulus. A zoom view of the right quadrant is provided in the right panel.

Finally, we compute equilibrium points and invariant manifolds of the dynamical system~\eqref{eqn:systmodel} in the rotating frame, using both, the original potential employed in the test particle simulation and the potential defined by the features we detect with our method. The original parameters we consider to generate the test particle simulation are detailed at the beginning of this section. As for the parameters detected by means of the presented methodology, we construct the potential as follows:

\begin{itemize}
\item A Ferrers bar, Eq.~\eqref{eqn:Ferrers}, characterised by potential $\phi_b$, with semi-major axis $a=3.786$ kpc, intermediate axis $b = 1.7233$ kpc, semi-minor axis $c = 0.3045$ kpc and mass $GM_b = 6.35 \%$ of the total mass.
\item A short bar or CORBE/DIRBE bulge, with potential $\phi_{sb}$ modeled as a Ferrers ellipsoid, as described in Eq.~\eqref{eqn:Ferrers}, featuring a semi-major axis $a = 3.2297$ kpc, semi-minor axes $b = 1.1125$ and $c = 0.323$, and mass $GM_{sb} = 4.32 \%$ of the total mass.
\item A Miyamoto-Nagai disc, following Eq.~\eqref{eqn:Miyamoto}, with potential $\phi_d$, and with parameters $A = 5.3178$ kpc, $B = 0.25$ kpc and $GM_d = 46.41\%$ of the total mass. 
\item A dark matter halo, Eq.~\eqref{eqn:halo}, characterised by a spherical potential $\phi_h$, extending to a maximum radius $R_{max} = 100$ kpc and $GM_h = 42.92\%$ of the total mass. 
\end{itemize}

The bar pattern speed is set to $\Omega = 0.053$~[u$_t$]$^{-1}$ ($\sim~25.93$~km/s/kpc).

Although the method does not explicitly identify the dark matter halo, its presence is inferred from the morphology of the arms. Under this bar pattern speed, the absence of a dark matter halo would result in more open arms~\citep[see][for a detailed analysis]{tesis}.

\begin{figure}
\centering
\includegraphics[width=0.7\textwidth]{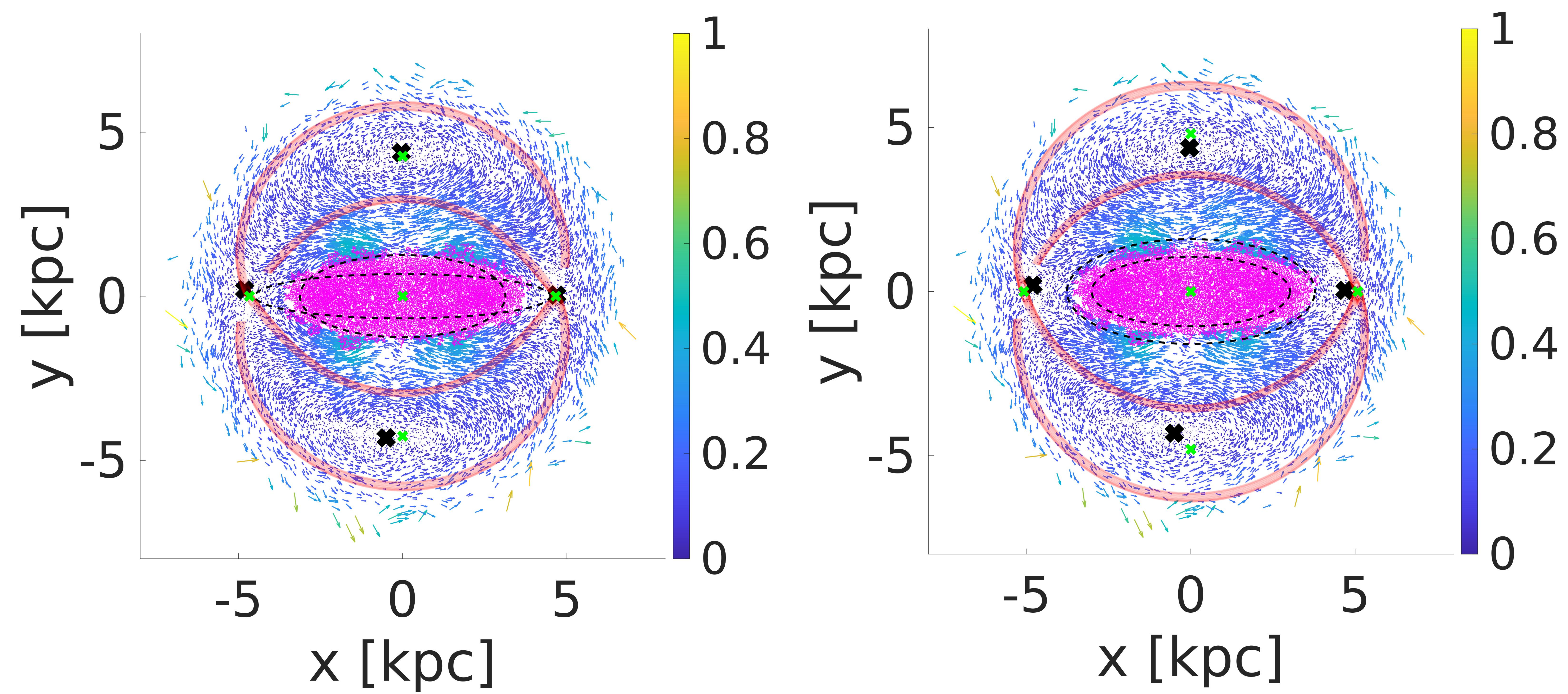}
\caption{Example from the test particle simulation. Invariant manifolds of the analytical model (in red) used for the test particle simulation and equilibrium points of the analytical model in green, superimposed onto Fig.~\ref{fig:test_barrabras}. The black dashed lines outline the position of the short and the large Ferrers bars. Left: The model is built using the data considered for the test particle simulation. Right: The model is built with the data extracted from the test particle simulation by means of the proposed method.} 
\label{fig:test_manif}
\end{figure}

Fig.~\ref{fig:test_manif} presents the equilibrium points (in green) and invariant manifolds (in red) of the analytical model~\eqref{eqn:systmodel}, representing a ring galaxy. The left panel illustrates the model components constructed from the data employed in creating the test particle simulation, while the right panel depicts the same components derived from the data obtained through the proposed methodology of analysis. In both cases, the invariant manifolds envelop the zero velocity curves, and the analytical equilibrium points closely coincide with the ones detected. The position of the short and large Ferrers bars is delineated by black dashed lines.

\subsection{Model calibration from an N-body simulation}

Subsequently, we investigate an N-body simulation crafted by \citet{Roca2013}, in which the structure of an isolated barred galaxy gradually emerges. The simulation covers a time span of 3.1~Gyr. The disc-to-halo mass ratio is set to an appropriate value to guarantee the formation of a robust bar and transient spiral arms over time. Employing a time step of 16~Myr, amounting to 6\% of the bar period, ensures a suitable resolution for inferring the bar pattern speed.

Our analysis focuses on snapshots from the simulation within the time interval [1.92, 2.40] Gyr, when the bar has  been clearly formed. 
In each snapshot, encompassing 5 million stars, the analysis follows the same procedure we use in the test particle case. Specifically, we examine 100 random batches, each containing $5\times10^5$ stars.
Again, we detect the bar features and critical points, tabulate the results, and subject them to normality tests.

\begin{figure}
\begin{center}
\includegraphics[width=0.32\textwidth]{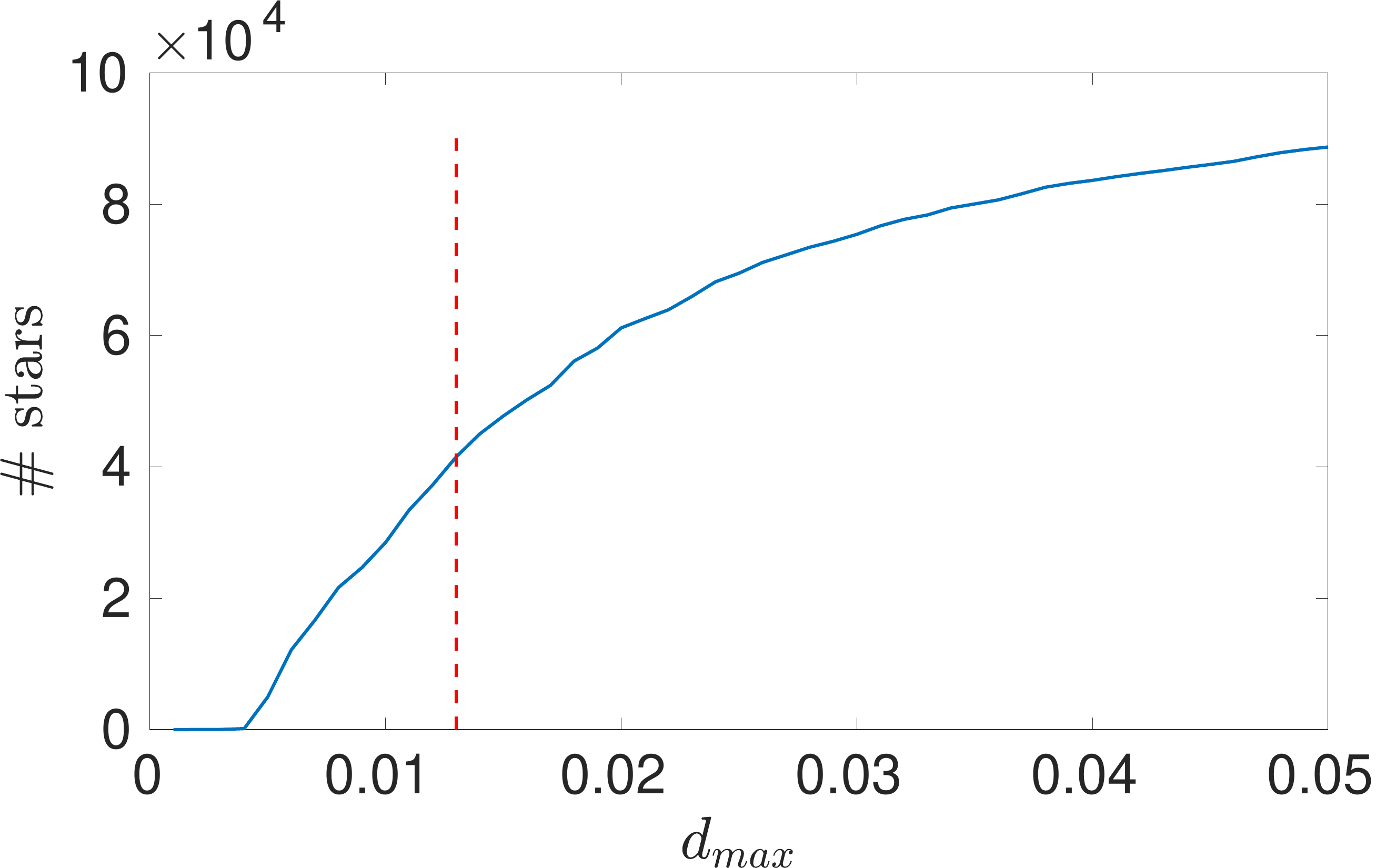}
\includegraphics[width=0.33\textwidth]{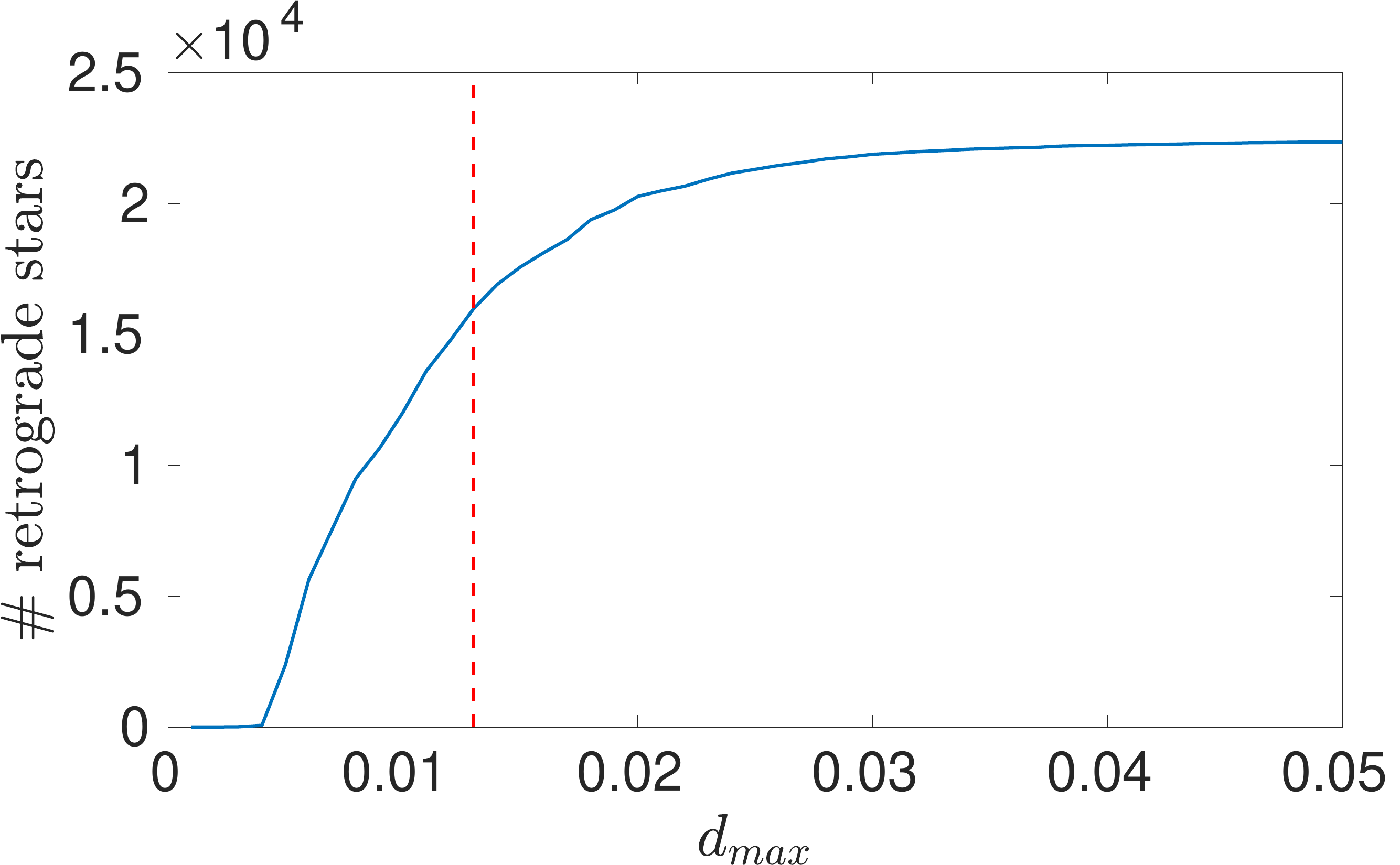}
\includegraphics[width=0.32\textwidth]{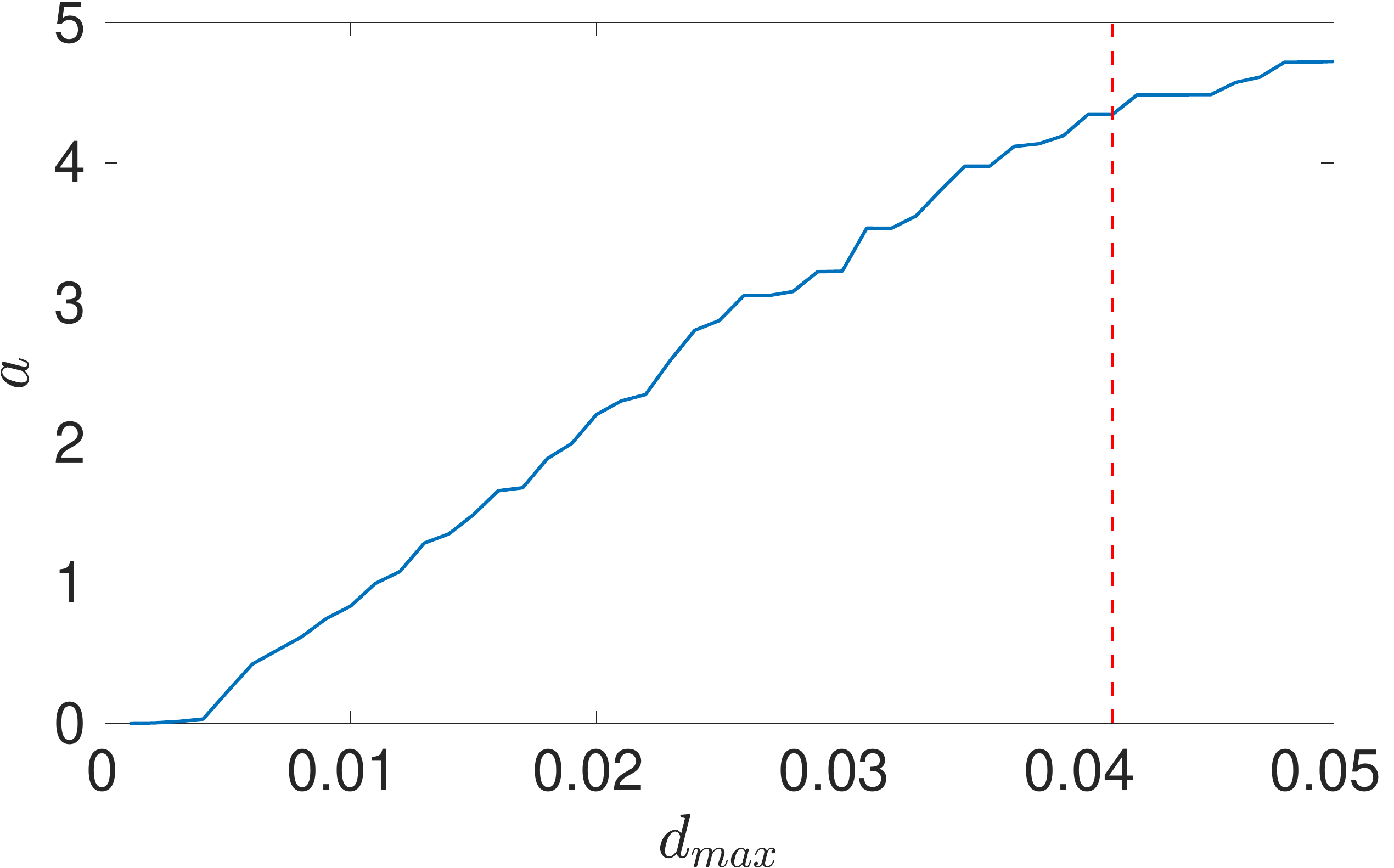} 
\end{center}
\caption{Sample batch 1 from the N-body simulation at time 2.24 Gyr : Growth of the connected component identified as the bar with respect to the persistent homology parameter $d_{max}$: (Left) in number of stars; (centre) in number of stars with a retrograde motion; (right) in main semiaxis length $a$. In the three cases, the point in the graph best fitting the shape $\Gamma$ is indicated in red.} \label{fig:dmaxnb}
\end{figure}

In this case, the distinction between the bar in a central position and the outlier stars with noncircular motion elsewhere in the galaxy is not as neat as in the previous test particle case. Figure~\ref{fig:dmaxnb} illustrates the fact showing the curves for a typical batch growing with $d_{max}$ for: number of stars with significant radial speed, stars with significant radial speed and retrograde motion, and finally, main semiaxis $a$ of the bar. The computation to detect the best fit point of the plot to the set square $\Gamma$ shape is required. In the computations we find that the curves with number of stars and of stars with a retrograde motion provide identical results, which give plausible results in the subsequent analysis and computations. The change of slope marking the limit of the bar is actually more pronounced for retrograde motion stars (see Fig.~\ref{fig:dmaxnb}). On the other hand, the bar semiaxis $a$ yields a completely different curve, where the fit with the $\Gamma$ shape to detect bar limit is much less robust, casting doubts on the usefulness of the growth of this parameter to estimate the bar size in galaxies with varied morphologies.

Table~\ref{taula:nb} presents the detected bar features for the snapshots of the simulation at times 1.92, 2.08, 2.24 and 2.40 Gyr. As in the test particle simulation, the method detects two bars: a shorter one that allows for the establishment of the pattern speed with high accuracy (again, for simplicity, the table only shows this short bar), and a larger one with semiaxes $a=3.0075$~kpc and $b=1.2331$kpc.

\begin{table*}
\resizebox{\textwidth}{!}{%
        \begin{tabular}{|c|c|c|c|c|c|c|c|}
                \hline
                Time [Gyr] & $a$ [kpc] & $b$ [kpc] & $\Omega$ [km s$^{-1}$ kpc$^{-1}$] & L1 & L2 & L4 & L5 \\ \hline
                $1.92$ & 1.5061 $\pm 0.2079$ & 0.7235 $\pm 0.0807$ & 20.5638 $\pm 0.7672$ & 
$\begin{pmatrix} 6.8300\\ -1.9521 \end{pmatrix} \pm \begin{pmatrix} 0.9052 \\ 1.0347 \end{pmatrix}$ &
$\begin{pmatrix} -7.3322 \\ 3.1117 \end{pmatrix} \pm \begin{pmatrix} 1.0744 \\ 0.5684 \end{pmatrix}$ &
$\begin{pmatrix} 7.6205 \\ 2.9050 \end{pmatrix} \pm \begin{pmatrix} 1.6123 \\ 1.5098 \end{pmatrix}$ &
$\begin{pmatrix} -4.5031 \\ -5.8087 \end{pmatrix} \pm \begin{pmatrix} 3.6652 \\ 0.9031 \end{pmatrix}$ \\ \hline
                $2.08$ & 1.1622 $\pm 0.1156$ & 0.6700 $\pm 0.0523$ & 22.4871 $\pm 1.0269$ & 
$\begin{pmatrix} 6.1071 \\ 0.6458 \end{pmatrix} \pm \begin{pmatrix} 0.9921 \\ 1.2688 \end{pmatrix}$ &
$\begin{pmatrix} -6.4692 \\ 0.3708 \end{pmatrix} \pm \begin{pmatrix} 0.9256 \\ 2.1557 \end{pmatrix}$ &
$\begin{pmatrix} -2.0181 \\ 5.4911 \end{pmatrix} \pm \begin{pmatrix} 4.6577 \\ 1.5346 \end{pmatrix}$ &
$\begin{pmatrix} -0.2271 \\ -5.7531 \end{pmatrix} \pm \begin{pmatrix} 5.0450 \\ 1.3317 \end{pmatrix}$ \\ \hline
                $2.24$  & 1.0705 $\pm 0.0652$ & 0.6547 $\pm 0.0313$ & 20.1674 $\pm 0.8845$ & 
$\begin{pmatrix} 7.5800 \\ 2.4705 \end{pmatrix} \pm \begin{pmatrix} 1.4049 \\ 1.9656 \end{pmatrix}$ &
$\begin{pmatrix} -6.0738 \\ -5.2179 \end{pmatrix} \pm \begin{pmatrix} 1.2660 \\ 2.3894 \end{pmatrix}$ &
$\begin{pmatrix} -1.0553 \\ 7.2317 \end{pmatrix} \pm \begin{pmatrix} 2.7584 \\ 1.7539 \end{pmatrix}$ &
$\begin{pmatrix} 3.0096 \\ -4.8924 \end{pmatrix} \pm \begin{pmatrix} 3.8779 \\ 3.4422 \end{pmatrix}$ \\ \hline
                $2.40$ & 0.8754 $\pm 0.0618$ & 0.6082 $\pm 0.0441$ & 19.7853 $\pm 0.9338$ & 
$\begin{pmatrix} 3.6794 \\ 7.9368 \end{pmatrix} \pm \begin{pmatrix} 1.9712 \\ 1.1564 \end{pmatrix}$ &
$\begin{pmatrix} -3.1627 \\ -7.2529 \end{pmatrix} \pm \begin{pmatrix} 2.4800 \\ 1.3037 \end{pmatrix}$ &
$\begin{pmatrix} -4.2751 \\ 6.9038 \end{pmatrix} \pm \begin{pmatrix} 2.6585 \\ 3.4017 \end{pmatrix}$ &
$\begin{pmatrix} 4.3758 \\ -5.1289 \end{pmatrix} \pm \begin{pmatrix} 3.0134 \\ 4.5509 \end{pmatrix}$ \\ \hline
        \end{tabular}}
        \caption{Detected bar features in the N-body simulation. All variables follow normal distribution. For equilibrium points, the standard deviation applies separately to each coordinate.}
        \label{taula:nb}
\end{table*}

We can see from Table~\ref{taula:nb} that the determination of the size of the bar is consistent across star samples and time snapshots, while the determination of the critical points of the system outside the bar is more uncertain. The estimates of the pattern speed $\Omega$ at each snapshot can be compared to the average pattern speed at which the detected bar has evolved between snapshots, which is respectively $21.6662,\,22.0548,\,22.3708$~km~s$^{-1}$~kpc$^{-1}$ between the four analyzed times. The average pattern speed of the bar over the time during which it is well defined has been estimated by the authors of \citet{Roca2013} as $21$~km~s$^{-1}$~kpc$^{-1}$.

\begin{figure}
\centering
\includegraphics[width=0.6\textwidth]{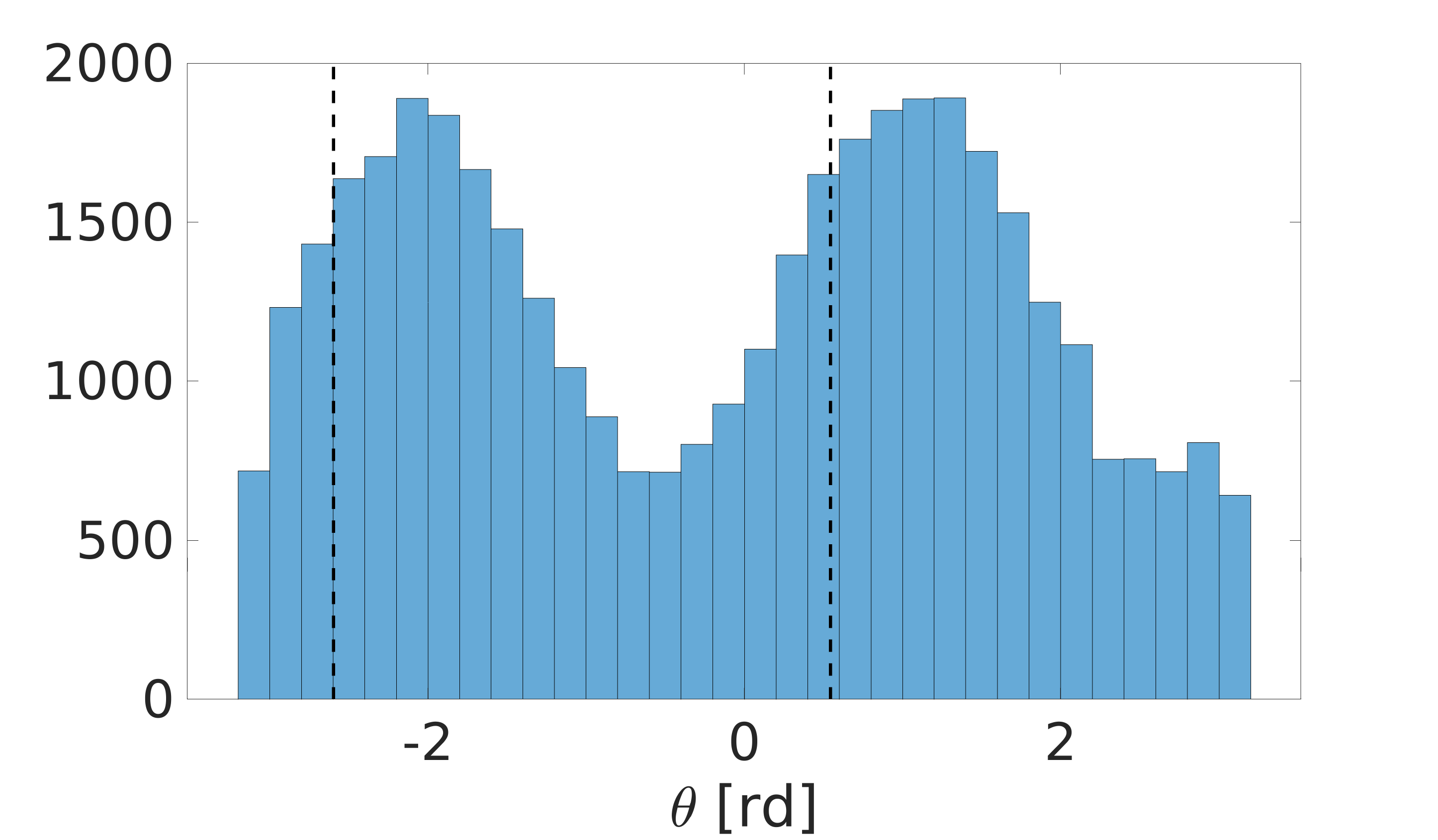}
\caption{Density distribution of the directions of motion of the particles constituting the bar in a sample population from time $t=2.08$ Gyr: the peak densities are close to the directions of the main axis of the bar as predicted in \citet{Athan1983}, with a small shift.} \label{fig:hist_dirs}
\end{figure}

It is noteworthy that our determination of stars constituting the bar agrees with the study of its components in \cite{Athan1983}. In this work, the authors argue that the central galactic bar is given consistency by a family of elongated periodic orbits around its centre, which follow the main bar axis for most of its run. Figure~\ref{fig:hist_dirs} compares the distribution of velocity directions for the particles that we have detected as bar constituents in a typical population at time $t=2.08$ Gyr with the direction of the main axis of the bar. We find that the velocity directions with peak density are those of the main axis with the addition of a small shift.

Probably related to the small shift in the peak direction distribution, we also review whether bar membership is a permanent or temporary condition for stars. For this test, again we randomly select sample batches of $5\times10^5$ stars, but this choice is held constant along all the snapshots. The results for a typical batch of $5\times10^5$ stars are shown in Table~\ref{taula:corrnb}. As in the previous simulation in section \ref{ss:pt} we see that belonging to the bar is a transitory state for most stars.

\begin{table}
        \centering
        \begin{tabular}{|c|c|c|c|}
                \hline
                Bars   & $B_1$ & $B_2$ & $B_3$ \\ \hline
                $B_1$  & 41534 & 13106 & 13002 \\ \hline
                $B_2$  &       & 40802 & 13162 \\ \hline
                $B_3$  &       &       & 39282 \\ \hline
        \end{tabular}
        \caption{Intersection table of (short) bars stars for batch 1 of 500.000 stars of the N-body simulation at times $t_1=2.08$ Gyr, $t_2=2.24$ Gyr, $t_3=2.40$ Gyr. $B_1,B_2,B_3$ denote the subset of batch stars which form the bar at times $t_1,t_2,t_3$ respectively. The table shows the number of stars in each set intersection $B_i \cap B_j$. The triple intersection $B_1 \cap B_2 \cap B_3$ contains 5031 stars.}
        \label{taula:corrnb}
\end{table}

\begin{figure}
\centering
\includegraphics[width=0.7\textwidth]{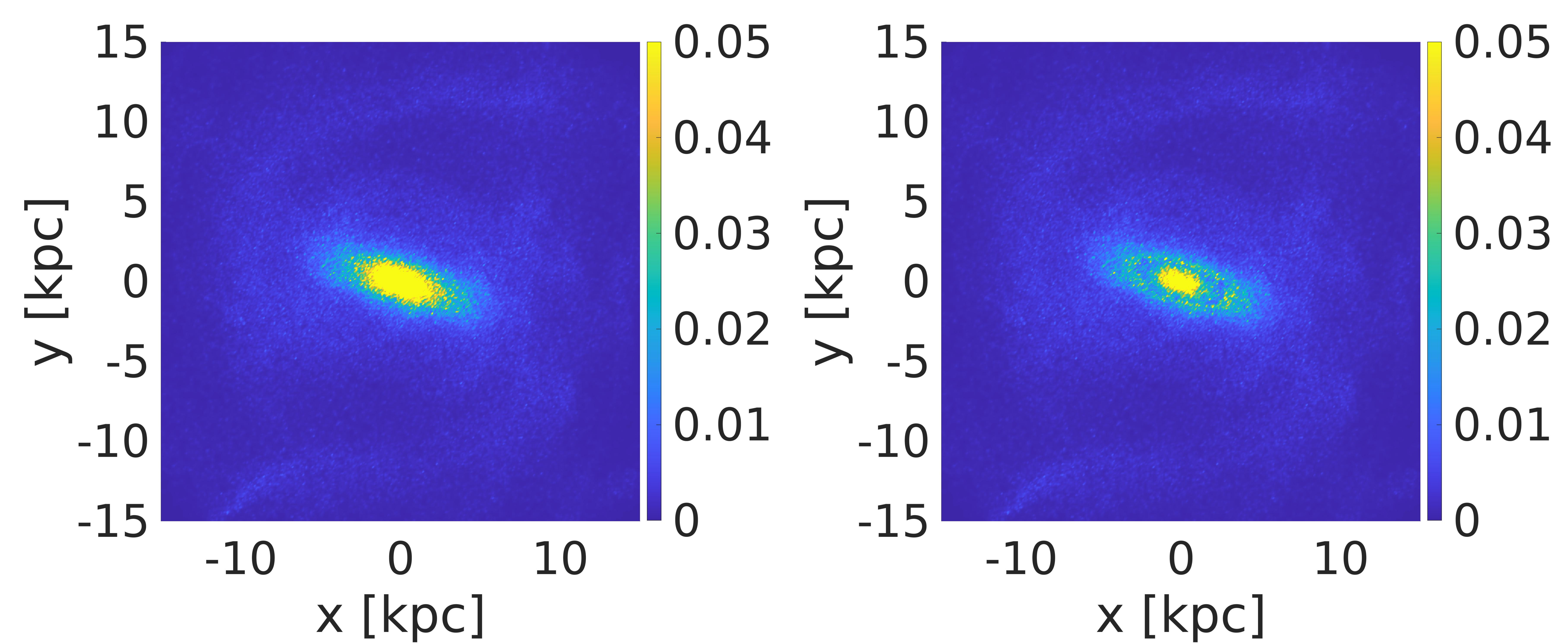}
\caption{Example from N-body simulation. Left: Density plot of all the particles in the simulation. Right: Density plot of all particles without the bar particles.} \label{fig:nb_denstot_y_disco}
\end{figure}

\begin{figure*}
\includegraphics[width=0.9\textwidth]{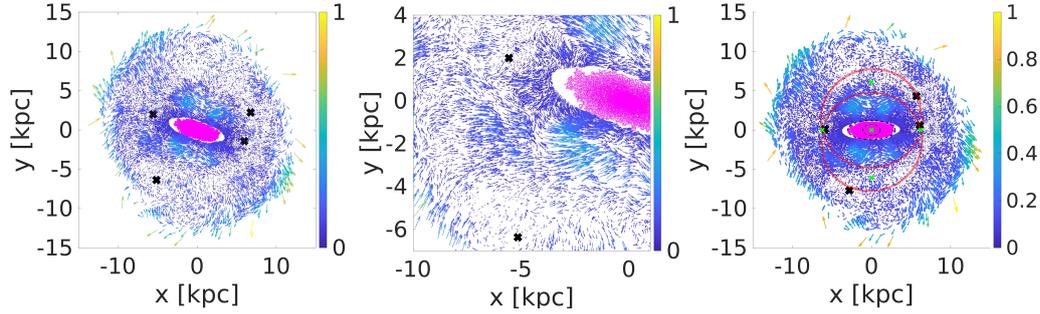}
\caption{Example from N-body simulation. Left: Bar stars (magenta); averaged velocity vectors of stars forming the arms (magnitude according to color bar), note that the stars in the inner arms move in a direction which is the opposite from those in the outer arms; estimated position for the equilibrium points (black). Center: Zoom-in view of a region. Right: Invariant manifolds of the analytical model (red) used in the test particle simulation superimposed to the left panel; equilibrium points of the analytical model in green; position of the short and large Ferrers bars outlined by black dashed lines.} 
\label{fig:nb_barrabras}
\end{figure*}

Figs.~\ref{fig:nb_denstot_y_disco}, \ref{fig:nb_barrabras} show the results of the proposed method for this N-body simulation. In the left panel of Fig.~\ref{fig:nb_denstot_y_disco} we see the density plot of the original data of the N-body simulation, while in the right panel of the same figure we see the density plot with the detected bar particles. The central region of the last panel contains a significant portion of particles.

As in the previous example, we compute the equilibrium points and the invariant manifolds of the dynamical system~\eqref{eqn:systmodel} in the rotating reference frame, using the potential we estimate with our method. The parameters we find for this potential are:
\begin{itemize}
\item A Ferrers bar, Eq.~\eqref{eqn:Ferrers} potential $\phi_b$, with semi-major axis $a=3$~kpc, intermediate axis $b = 1.23$~kpc, semi-minor axis $c = 0.3045$~kpc and mass $GM_b = 20.12 \%$ of the total mass.
\item A short bar or CORBE/DIRBE bulge, potential $\phi_{sb}$ modelled as a Ferrers ellipsoid Eq.~\eqref{eqn:Ferrers}, with a semi-major axis $a = 1.1$~kpc, semi-minor axes $b = 0.7$~kpc and $c = 0.3045$~kpc, and mass $GM_{sb} = 4.08 \%$ of the total mass.
\item A Miyamoto-Nagai disc, Eq.~\eqref{eqn:Miyamoto} potential $\phi_d$, with parameters $A = 15.32$ kpc, $B = 1.25$ kpc and $GM_d = 75.8\%$ of the total mass. 
\end{itemize}
Taking into account that $G(M_d+M_b+M_{sb}) = 1$. The bar pattern speed is set to $\Omega = 0.041$~[u$_t$]$^{-1}$ ($\sim 20.06$~km/s/kpc).

As we illustrate in Fig.~\ref{fig:nb_barrabras} using the non-inertial (rotating) reference system,
the arms are highlighted by the larger size of the averaged velocity vectors of stars 
 and the estimated position for the equilibrium points are marked in black.
The regions where the modulus of the vectors is close to zero delineate also the Hill's region of the system. 
The central panel of Fig.~\ref{fig:nb_barrabras} displays a zoomed-in view of a quadrant from the left panel. Finally, in the right panel, we can see the invariant manifolds (red) and the equilibrium points of the analytical model (green) revealing a spiral galaxy. These are computed using the potential components calibrated from the data obtained by the methodology.

\section{Conclusions}
\label{sec:disc} 

The objective of this study is to introduce a novel procedure to identify key features and the main building blocks of galaxies, such as bars, discs and arms, using a single snapshot and without relying on particle density.

Our work confirms prior findings that the bar and arms in a barred galaxy are not fixed sets of stars following a particular motion. We find that they can be seen as perturbations in the radial (with respect to the galactic center) velocity of the galaxy stars. These perturbations rotate in the galactic disc, and many of their component stars belong only transiently to the bar or the arms.

The proposed approach is applicable to a barred galaxy in which the instantaneous positions and velocities of a statistically representative sample of stars are known. Our procedure takes as input a table of positions and velocities of the sample of galaxy stars at a fixed time. In the case in which the line-of-sight component of position and velocity is known with a greater error, only the components of position and velocity in the galactic plane are an essential requirement. We employ the $k$-means clustering algorithm to categorize particles into disc and non-disc populations. Subsequently, we identify bars using persistent homology techniques, which enable the determination of bar pattern speeds and facilitate the transformation of coordinates from an inertial to a rotating reference frame. Within the rotating frame, we estimate the equilibrium points and the arms of the galaxy. The method provides reliable results for detecting the building blocks and bar pattern speed of the galaxy. Additionally, it estimates bar and arm characteristics even if the stellar masses or the time duration for which stars remain bound to these structures are unknown. This approach is also applicable to simulations, N-body or otherwise, of galaxies.

The validation of this approach is conducted using both test particle and N-body simulations. In spite of the utilisation of a single snapshot as starting data, and the fact that the bar and arms are travelling perturbations with transient membership, the accuracy of the method is established by comparing results across multiple snapshots.
The test particle simulation demonstrates the high accuracy of the method in identifying key features, including two bars and the corresponding pattern speed, as well as locating equilibrium points and arms with minimal error. Results from the N-body simulation exhibit larger errors due to inherent simulation complexities; however, the method remains acceptable in terms of error margins across all features. Notably, in this simulation the method can identify two bars, with their corresponding pattern speeds and approximate equilibrium points.

In both simulations, we consider the detected bars and disc sizes, short bar pattern speed and bars and disc masses (under the statistical assumption that all particles in the simulation have equal mass) to build a potential model of the galaxy. This model consists in a simple Miyamoto-Nagai disc and two Ferrers bars. 
In this dynamical system we study the equilibrium points and the invariant manifolds that emanate from the periodic orbits around the unstable equilibrium points at the ends of the bar.  

The invariant manifold theory is based on the fact that the equilibrium points at the ends of the bar are surrounded by libration point orbits, which are normally hyperbolic with stable and unstable manifolds associated to them. The unstable manifolds emanate from the periodic Lyapunov orbits in the form of channels that sketch the galactic bar structure. Besides, the stretching property of unstable manifolds (\cite{LCS2018}) compresses also exterior matter towards them, making the arms more visible.
This study additionally explores and enhances the application of invariant manifold theory to understand arm formations, overlapping theoretical structures with the results obtained from the proposed method.
There exists a close alignment when we superimpose these structures, obtained from the theoretical model, on top the particle data from the simulations.  

The authors expect that this methodology will enable the work with sampled real data, quantifying local structures in the short-term for near galaxies such as the Large Magellanic Cloud.

\section{Acknowledgements}
P.S.M. thanks the Spanish Ministry of Economy grants PID2020-117066GB-I00 and PID2021-123968NB-I00.
J.A. thanks the Spanish Ministry of Economy grants PID2023-146936NB-I00 and the AGAUR grant 2021 SGR 00603.
J.J.M. thanks MINECO-FEDER for the grant PID2021-123968NB-I00.

\bibliographystyle{elsarticle-harv} 

\bibliography{biblio} 

\begin{thebibliography}{46}
\expandafter\ifx\csname natexlab\endcsname\relax\def\natexlab#1{#1}\fi
\providecommand{\url}[1]{\texttt{#1}}
\providecommand{\href}[2]{#2}
\providecommand{\path}[1]{#1}
\providecommand{\DOIprefix}{doi:}
\providecommand{\ArXivprefix}{arXiv:}
\providecommand{\URLprefix}{URL: }
\providecommand{\Pubmedprefix}{pmid:}
\providecommand{\doi}[1]{\href{http://dx.doi.org/#1}{\path{#1}}}
\providecommand{\Pubmed}[1]{\href{pmid:#1}{\path{#1}}}
\providecommand{\bibinfo}[2]{#2}
\ifx\xfnm\relax \def\xfnm[#1]{\unskip,\space#1}\fi
\bibitem[{Allen and Santillan(1991)}]{Allen1991}
\bibinfo{author}{Allen, C.}, \bibinfo{author}{Santillan, A.},
  \bibinfo{year}{1991}.
\newblock \bibinfo{title}{An improved model of the galactic mass distribution
  for orbit computations}.
\newblock \bibinfo{journal}{Revista Mexicana de Astronomia y Astrofisica}
  \bibinfo{volume}{22}, \bibinfo{pages}{255--263}.
\bibitem[{Athanassoula(2003)}]{Athan2003}
\bibinfo{author}{Athanassoula, E.}, \bibinfo{year}{2003}.
\newblock \bibinfo{title}{What determines the strength and the slowdown rate of
  bars?}
\newblock \bibinfo{journal}{MNRAS} \bibinfo{volume}{341},
  \bibinfo{pages}{1179--1198}.
\bibitem[{Athanassoula et~al.(1983)Athanassoula, Bienayme, Martinet and
  Pfenniger}]{Athan1983}
\bibinfo{author}{Athanassoula, E.}, \bibinfo{author}{Bienayme, O.},
  \bibinfo{author}{Martinet, L.}, \bibinfo{author}{Pfenniger, D.},
  \bibinfo{year}{1983}.
\newblock \bibinfo{title}{Orbits as building blocks of a barred galaxy model}.
\newblock \bibinfo{journal}{A\&A} \bibinfo{volume}{127},
  \bibinfo{pages}{349--360}.
\bibitem[{Brown et~al.(2021)Brown, Vallenari, Prusti, De~Bruijne, Babusiaux,
  Biermann, Creevey, Evans, Eyer, Hutton et~al.}]{Gaia2021}
\bibinfo{author}{Brown, A.G.}, \bibinfo{author}{Vallenari, A.},
  \bibinfo{author}{Prusti, T.}, \bibinfo{author}{De~Bruijne, J.H.},
  \bibinfo{author}{Babusiaux, C.}, \bibinfo{author}{Biermann, M.},
  \bibinfo{author}{Creevey, O.L.}, \bibinfo{author}{Evans, D.W.},
  \bibinfo{author}{Eyer, L.}, \bibinfo{author}{Hutton, A.}, et~al.,
  \bibinfo{year}{2021}.
\newblock \bibinfo{title}{Gaia early data release 3-summary of the contents and
  survey properties}.
\newblock \bibinfo{journal}{A\&A} \bibinfo{volume}{649}, \bibinfo{pages}{A1}.
\bibitem[{Canalias and Masdemont(2006)}]{CanaliasMasdemont2006}
\bibinfo{author}{Canalias, E.}, \bibinfo{author}{Masdemont, J.J.},
  \bibinfo{year}{2006}.
\newblock \bibinfo{title}{Homoclinic and heteroclinic transfer trajectories
  between planar lyapunov orbits in the sun-earth and earth-moon systems}.
\newblock \bibinfo{journal}{Discrete and continuous dynamical systems}
  \bibinfo{volume}{14}, \bibinfo{pages}{261}.
\bibitem[{Chiba et~al.(2021)Chiba, Friske and Sch{\"o}nrich}]{ChibaFS2021}
\bibinfo{author}{Chiba, R.}, \bibinfo{author}{Friske, J.K.},
  \bibinfo{author}{Sch{\"o}nrich, R.}, \bibinfo{year}{2021}.
\newblock \bibinfo{title}{Resonance sweeping by a decelerating galactic bar}.
\newblock \bibinfo{journal}{MNRAS} \bibinfo{volume}{500},
  \bibinfo{pages}{4710--4729}.
\bibitem[{Debattista et~al.(2017)Debattista, Ness, Gonzalez, Freeman, Zoccali
  and Minniti}]{DNGFZM2017}
\bibinfo{author}{Debattista, V.P.}, \bibinfo{author}{Ness, M.},
  \bibinfo{author}{Gonzalez, O.A.}, \bibinfo{author}{Freeman, K.},
  \bibinfo{author}{Zoccali, M.}, \bibinfo{author}{Minniti, D.},
  \bibinfo{year}{2017}.
\newblock \bibinfo{title}{Separation of stellar populations by an evolving bar:
  implications for the bulge of the milky way}.
\newblock \bibinfo{journal}{MNRAS} \bibinfo{volume}{469},
  \bibinfo{pages}{1587--1611}.
\bibitem[{Dehnen et~al.(2023)Dehnen, Semczuk and Sch{\"o}nrich}]{Dehnen2023}
\bibinfo{author}{Dehnen, W.}, \bibinfo{author}{Semczuk, M.},
  \bibinfo{author}{Sch{\"o}nrich, R.}, \bibinfo{year}{2023}.
\newblock \bibinfo{title}{Measuring bar pattern speeds from single simulation
  snapshots}.
\newblock \bibinfo{journal}{MNRAS} \bibinfo{volume}{518},
  \bibinfo{pages}{2712--2718}.
\bibitem[{Duda et~al.(2001)Duda, Hart and Stork}]{DHS2001}
\bibinfo{author}{Duda, R.O.}, \bibinfo{author}{Hart, P.E.},
  \bibinfo{author}{Stork, D.G.}, \bibinfo{year}{2001}.
\newblock \bibinfo{title}{Pattern classification}.
\newblock \bibinfo{journal}{second edition john wiley \& sons, New York}
  \bibinfo{volume}{58}, \bibinfo{pages}{16}.
\bibitem[{Edelsbrunner and Harer(2022)}]{EH2022}
\bibinfo{author}{Edelsbrunner, H.}, \bibinfo{author}{Harer, J.L.},
  \bibinfo{year}{2022}.
\newblock \bibinfo{title}{Computational topology: an introduction}.
\newblock \bibinfo{publisher}{American Mathematical Society}.
\bibitem[{Fantino et~al.(2023)Fantino, Burhani, Flores, Alessi, Solano and
  Sanjurjo-Rivo}]{Fantino2023}
\bibinfo{author}{Fantino, E.}, \bibinfo{author}{Burhani, B.M.},
  \bibinfo{author}{Flores, R.}, \bibinfo{author}{Alessi, E.M.},
  \bibinfo{author}{Solano, F.}, \bibinfo{author}{Sanjurjo-Rivo, M.},
  \bibinfo{year}{2023}.
\newblock \bibinfo{title}{End-to-end trajectory concept for close exploration
  of saturn’s inner large moons}.
\newblock \bibinfo{journal}{Communications in Nonlinear Science and Numerical
  Simulation} \bibinfo{volume}{126}, \bibinfo{pages}{107458}.
\bibitem[{Ferrers(1877)}]{Ferrers}
\bibinfo{author}{Ferrers, N.}, \bibinfo{year}{1877}.
\newblock \bibinfo{title}{On the potentials, ellipsoids, ellipsoidal shells,
  elliptic laminae, and elliptics rings, of variable densities}.
\newblock \bibinfo{journal}{The Quarterly Journal of Pure and Applied
  Mathematics} \bibinfo{volume}{14}, \bibinfo{pages}{1--22}.
\bibitem[{Fragkoudi et~al.(2021)Fragkoudi, Grand, Pakmor, Springel, White,
  Marinacci, Gomez and Navarro}]{Fragkoudi2021}
\bibinfo{author}{Fragkoudi, F.}, \bibinfo{author}{Grand, R.J.},
  \bibinfo{author}{Pakmor, R.}, \bibinfo{author}{Springel, V.},
  \bibinfo{author}{White, S.D.}, \bibinfo{author}{Marinacci, F.},
  \bibinfo{author}{Gomez, F.A.}, \bibinfo{author}{Navarro, J.F.},
  \bibinfo{year}{2021}.
\newblock \bibinfo{title}{Revisiting the tension between fast bars and the
  $\lambda$cdm paradigm}.
\newblock \bibinfo{journal}{A\&A} \bibinfo{volume}{650}, \bibinfo{pages}{L16}.
\bibitem[{Gidea and Masdemont(2007)}]{GideaMasdem2007}
\bibinfo{author}{Gidea, M.}, \bibinfo{author}{Masdemont, J.J.},
  \bibinfo{year}{2007}.
\newblock \bibinfo{title}{Geometry of homoclinic connections in a planar
  circular restricted three-body problem}.
\newblock \bibinfo{journal}{International journal of bifurcation and chaos}
  \bibinfo{volume}{17}, \bibinfo{pages}{1151--1169}.
\bibitem[{G{\'o}mez et~al.(2004)G{\'o}mez, Koon, Lo, Marsden, Masdemont and
  Ross}]{GomezKoon2004}
\bibinfo{author}{G{\'o}mez, G.}, \bibinfo{author}{Koon, W.S.},
  \bibinfo{author}{Lo, M.W.}, \bibinfo{author}{Marsden, J.E.},
  \bibinfo{author}{Masdemont, J.}, \bibinfo{author}{Ross, S.D.},
  \bibinfo{year}{2004}.
\newblock \bibinfo{title}{Connecting orbits and invariant manifolds in the
  spatial restricted three-body problem}.
\newblock \bibinfo{journal}{Nonlinearity} \bibinfo{volume}{17},
  \bibinfo{pages}{1571}.
\bibitem[{Henry and Scheeres(2024)}]{HenryScheeres2024}
\bibinfo{author}{Henry, D.B.}, \bibinfo{author}{Scheeres, D.J.},
  \bibinfo{year}{2024}.
\newblock \bibinfo{title}{Fully numerical computation of heteroclinic
  connection families in the spatial three-body problem}.
\newblock \bibinfo{journal}{Communications in Nonlinear Science and Numerical
  Simulation} \bibinfo{volume}{130}, \bibinfo{pages}{107780}.
\bibitem[{{Jim{\'e}nez-Arranz} et~al.(2024){Jim{\'e}nez-Arranz}, {Chemin},
  {Romero-G{\'o}mez}, {Luri}, {Adamczyk}, {Castro-Ginard}, {Roca-F{\`a}brega},
  {McMillan} and {Cioni}}]{JimenezArranz2024}
\bibinfo{author}{{Jim{\'e}nez-Arranz}, {\'O}.}, \bibinfo{author}{{Chemin}, L.},
  \bibinfo{author}{{Romero-G{\'o}mez}, M.}, \bibinfo{author}{{Luri}, X.},
  \bibinfo{author}{{Adamczyk}, P.}, \bibinfo{author}{{Castro-Ginard}, A.},
  \bibinfo{author}{{Roca-F{\`a}brega}, S.}, \bibinfo{author}{{McMillan}, P.J.},
  \bibinfo{author}{{Cioni}, M.R.L.}, \bibinfo{year}{2024}.
\newblock \bibinfo{title}{{The bar pattern speed of the Large Magellanic
  Cloud}}.
\newblock \bibinfo{journal}{A\&A} \bibinfo{volume}{683}, \bibinfo{pages}{A102}.
\bibitem[{{Kalda} et~al.(2024){Kalda}, {Green} and {Ghosh}}]{Kalda2024}
\bibinfo{author}{{Kalda}, T.}, \bibinfo{author}{{Green}, G.M.},
  \bibinfo{author}{{Ghosh}, S.}, \bibinfo{year}{2024}.
\newblock \bibinfo{title}{{Recovering the gravitational potential in a rotating
  frame: Deep Potential applied to a simulated barred galaxy}}.
\newblock \bibinfo{journal}{MNRAS} \bibinfo{volume}{527},
  \bibinfo{pages}{12284--12297}.
\bibitem[{Katsanikas and Patsis(2022)}]{Katsanikas2022}
\bibinfo{author}{Katsanikas, M.}, \bibinfo{author}{Patsis, P.},
  \bibinfo{year}{2022}.
\newblock \bibinfo{title}{The phase space structure in the vicinity of vertical
  lyapunov orbits around l 1, 2 in a barred galaxy model}.
\newblock \bibinfo{journal}{MNRAS} \bibinfo{volume}{516},
  \bibinfo{pages}{5232--5243}.
\bibitem[{Koon et~al.(2000)Koon, Lo, Marsden and Ross}]{KoonLo2000}
\bibinfo{author}{Koon, W.S.}, \bibinfo{author}{Lo, M.W.},
  \bibinfo{author}{Marsden, J.E.}, \bibinfo{author}{Ross, S.D.},
  \bibinfo{year}{2000}.
\newblock \bibinfo{title}{Heteroclinic connections between periodic orbits and
  resonance transitions in celestial mechanics}.
\newblock \bibinfo{journal}{Chaos: An Interdisciplinary Journal of Nonlinear
  Science} \bibinfo{volume}{10}, \bibinfo{pages}{427--469}.
\bibitem[{Marchuk et~al.(2024)Marchuk, Mosenkov, Chugunov, Kostiuk, Skryabina
  and Reshetnikov}]{Marchuk2024}
\bibinfo{author}{Marchuk, A.A.}, \bibinfo{author}{Mosenkov, A.V.},
  \bibinfo{author}{Chugunov, I.V.}, \bibinfo{author}{Kostiuk, V.S.},
  \bibinfo{author}{Skryabina, M.N.}, \bibinfo{author}{Reshetnikov, V.P.},
  \bibinfo{year}{2024}.
\newblock \bibinfo{title}{A new, purely photometric method for determination of
  resonance locations in spiral galaxies}.
\newblock \bibinfo{journal}{MNRAS} \bibinfo{volume}{527},
  \bibinfo{pages}{L66--L70}.
\bibitem[{{Miyamoto} and {Nagai}(1975)}]{Miyamoto}
\bibinfo{author}{{Miyamoto}, M.}, \bibinfo{author}{{Nagai}, R.},
  \bibinfo{year}{1975}.
\newblock \bibinfo{title}{{Three-dimensional models for the distribution of
  mass in galaxies.}}
\newblock \bibinfo{journal}{PASJ} \bibinfo{volume}{27},
  \bibinfo{pages}{533--543}.
\bibitem[{{Moges} et~al.(2024){Moges}, {Katsanikas}, {Patsis}, {Hillebrand} and
  {Skokos}}]{Moges2024}
\bibinfo{author}{{Moges}, H.T.}, \bibinfo{author}{{Katsanikas}, M.},
  \bibinfo{author}{{Patsis}, P.A.}, \bibinfo{author}{{Hillebrand}, M.},
  \bibinfo{author}{{Skokos}, C.}, \bibinfo{year}{2024}.
\newblock \bibinfo{title}{{The Evolution of the Phase Space Structure Along
  Pitchfork and Period-Doubling Bifurcations in a 3D-Galactic Bar Potential}}.
\newblock \bibinfo{journal}{International Journal of Bifurcation and Chaos}
  \bibinfo{volume}{34}, \bibinfo{pages}{2430013--360}.
\bibitem[{Patsis and Athanassoula(2019)}]{Patsis2019}
\bibinfo{author}{Patsis, P.}, \bibinfo{author}{Athanassoula, E.},
  \bibinfo{year}{2019}.
\newblock \bibinfo{title}{The orbital content of bars: the origin of
  ‘non-x1-tree’, bar-supporting orbits}.
\newblock \bibinfo{journal}{MNRAS} \bibinfo{volume}{490},
  \bibinfo{pages}{2740--2759}.
\bibitem[{Peschken and {\L}okas(2019)}]{Peschken2019}
\bibinfo{author}{Peschken, N.}, \bibinfo{author}{{\L}okas, E.L.},
  \bibinfo{year}{2019}.
\newblock \bibinfo{title}{Tidally induced bars in illustris galaxies}.
\newblock \bibinfo{journal}{MNRAS} \bibinfo{volume}{483},
  \bibinfo{pages}{2721--2735}.
\bibitem[{{Petersen} et~al.(2024){Petersen}, {Weinberg} and
  {Katz}}]{Petersen2024}
\bibinfo{author}{{Petersen}, M.S.}, \bibinfo{author}{{Weinberg}, M.D.},
  \bibinfo{author}{{Katz}, N.}, \bibinfo{year}{2024}.
\newblock \bibinfo{title}{{Measuring the dynamical length of galactic bars}}.
\newblock \bibinfo{journal}{MNRAS} \bibinfo{volume}{531},
  \bibinfo{pages}{751--763}.
\bibitem[{Pfenniger(1984)}]{Pfenn}
\bibinfo{author}{Pfenniger, D.}, \bibinfo{year}{1984}.
\newblock \bibinfo{title}{The 3d dynamics of barred galaxies}.
\newblock \bibinfo{journal}{A\&A} \bibinfo{volume}{134},
  \bibinfo{pages}{373--386}.
\bibitem[{Pfenniger et~al.(2023)Pfenniger, Saha and Wu}]{Pfenniger2023}
\bibinfo{author}{Pfenniger, D.}, \bibinfo{author}{Saha, K.},
  \bibinfo{author}{Wu, Y.T.}, \bibinfo{year}{2023}.
\newblock \bibinfo{title}{Five methods for determining pattern speeds in
  galaxies-i. methods}.
\newblock \bibinfo{journal}{A\&A} \bibinfo{volume}{673}, \bibinfo{pages}{A36}.
\bibitem[{Roca-F{\`a}brega et~al.(2013)Roca-F{\`a}brega, Valenzuela, Figueras,
  Romero-G{\'o}mez, Vel{\'a}zquez, Antoja and Pichardo}]{Roca2013}
\bibinfo{author}{Roca-F{\`a}brega, S.}, \bibinfo{author}{Valenzuela, O.},
  \bibinfo{author}{Figueras, F.}, \bibinfo{author}{Romero-G{\'o}mez, M.},
  \bibinfo{author}{Vel{\'a}zquez, H.}, \bibinfo{author}{Antoja, T.},
  \bibinfo{author}{Pichardo, B.}, \bibinfo{year}{2013}.
\newblock \bibinfo{title}{On galaxy spiral arms’ nature as revealed by
  rotation frequencies}.
\newblock \bibinfo{journal}{MNRAS} \bibinfo{volume}{432},
  \bibinfo{pages}{2878--2885}.
\bibitem[{{Romero-G{\'o}mez} et~al.(2011){Romero-G{\'o}mez}, {Athanassoula},
  {Antoja} and {Figueras}}]{Romero2011}
\bibinfo{author}{{Romero-G{\'o}mez}, M.}, \bibinfo{author}{{Athanassoula}, E.},
  \bibinfo{author}{{Antoja}, T.}, \bibinfo{author}{{Figueras}, F.},
  \bibinfo{year}{2011}.
\newblock \bibinfo{title}{{Modelling the inner disc of the Milky Way with
  manifolds - I. A first step}}.
\newblock \bibinfo{journal}{MNRAS} \bibinfo{volume}{418},
  \bibinfo{pages}{1176--1193}.
\bibitem[{Romero-G{\'o}mez et~al.(2007)Romero-G{\'o}mez, Athanassoula,
  Masdemont and Garc{\'\i}a-G{\'o}mez}]{Romero2}
\bibinfo{author}{Romero-G{\'o}mez, M.}, \bibinfo{author}{Athanassoula, E.},
  \bibinfo{author}{Masdemont, J.}, \bibinfo{author}{Garc{\'\i}a-G{\'o}mez, C.},
  \bibinfo{year}{2007}.
\newblock \bibinfo{title}{The formation of spiral arms and rings in barred
  galaxies}.
\newblock \bibinfo{journal}{A\&A} \bibinfo{volume}{472},
  \bibinfo{pages}{63--75}.
\bibitem[{Romero-G\'omez et~al.(2015)Romero-G\'omez, Figueras, Antoja, Abedi
  and Aguilar}]{Rom15}
\bibinfo{author}{Romero-G\'omez, M.}, \bibinfo{author}{Figueras, F.},
  \bibinfo{author}{Antoja, T.}, \bibinfo{author}{Abedi, H.},
  \bibinfo{author}{Aguilar, L.}, \bibinfo{year}{2015}.
\newblock \bibinfo{title}{The analysis of realistic stellar gaia mock
  catalogues--i. red clump stars as tracers of the central bar}.
\newblock \bibinfo{journal}{MNRAS} \bibinfo{volume}{447},
  \bibinfo{pages}{218--233}.
\bibitem[{Romero-G\'omez et~al.(2006)Romero-G\'omez, Masdemont, Athanassoula
  and Garc{\'\i}a-G{\'o}mez}]{Romero1}
\bibinfo{author}{Romero-G\'omez, M.}, \bibinfo{author}{Masdemont, J.},
  \bibinfo{author}{Athanassoula, E.}, \bibinfo{author}{Garc{\'\i}a-G{\'o}mez,
  C.}, \bibinfo{year}{2006}.
\newblock \bibinfo{title}{The origin of rr1 ring structures in barred
  galaxies}.
\newblock \bibinfo{journal}{A\&A} \bibinfo{volume}{453},
  \bibinfo{pages}{39--45}.
\bibitem[{Romero-G{\'o}mez et~al.(2009)Romero-G{\'o}mez, Masdemont,
  Garc{\'\i}a-G{\'o}mez and Athanassoula}]{Rom09}
\bibinfo{author}{Romero-G{\'o}mez, M.}, \bibinfo{author}{Masdemont, J.},
  \bibinfo{author}{Garc{\'\i}a-G{\'o}mez, C.}, \bibinfo{author}{Athanassoula,
  E.}, \bibinfo{year}{2009}.
\newblock \bibinfo{title}{The role of the unstable equilibrium points in the
  transfer of matter in galactic potentials}.
\newblock \bibinfo{journal}{Communications in Nonlinear Science and Numerical
  Simulation} \bibinfo{volume}{14}, \bibinfo{pages}{4123--4138}.
\bibitem[{{S{\'a}nchez Mart{\'\i}n}(2015)}]{tesis}
\bibinfo{author}{{S{\'a}nchez Mart{\'\i}n}, P.}, \bibinfo{year}{2015}.
\newblock \bibinfo{title}{{Application of dynamical system methods to galactic
  dynamics : from warps to double bars}}.
\newblock Ph.D. thesis. Universitat Polit\`ecnica de Catalunya, Spain.
\bibitem[{S{\'a}nchez-Mart{\'\i}n et~al.(2023)S{\'a}nchez-Mart{\'\i}n,
  Garc{\'\i}a-G{\'o}mez, Masdemont and Romero-G{\'o}mez}]{Asymmetry}
\bibinfo{author}{S{\'a}nchez-Mart{\'\i}n, P.},
  \bibinfo{author}{Garc{\'\i}a-G{\'o}mez, C.}, \bibinfo{author}{Masdemont,
  J.J.}, \bibinfo{author}{Romero-G{\'o}mez, M.}, \bibinfo{year}{2023}.
\newblock \bibinfo{title}{Formation of asymmetric arms in barred galaxies}.
\newblock \bibinfo{journal}{MNRAS} \bibinfo{volume}{520},
  \bibinfo{pages}{3909--3915}.
\bibitem[{S\'anchez-Mart\'in et~al.(2018)S\'anchez-Mart\'in, Masdemont and
  Romero-G\'omez}]{LCS2018}
\bibinfo{author}{S\'anchez-Mart\'in, P.}, \bibinfo{author}{Masdemont, J.},
  \bibinfo{author}{Romero-G\'omez, M.}, \bibinfo{year}{2018}.
\newblock \bibinfo{title}{From manifolds to lagrangian coherent structures in
  galactic bar models}.
\newblock \bibinfo{journal}{A\&A} \bibinfo{volume}{618}, \bibinfo{pages}{A72}.
\bibitem[{S\'anchez-Mart\'in et~al.(2016)S\'anchez-Mart\'in, Romero-G\'omez and
  Masdemont}]{Warps}
\bibinfo{author}{S\'anchez-Mart\'in, P.}, \bibinfo{author}{Romero-G\'omez, M.},
  \bibinfo{author}{Masdemont, J.J.}, \bibinfo{year}{2016}.
\newblock \bibinfo{title}{Warp evidence in precessing galactic bar models}.
\newblock \bibinfo{journal}{A\&A} \bibinfo{volume}{588}, \bibinfo{pages}{A76}.
\bibitem[{Sellwood and Athanassoula(1986)}]{SellwoodAthan1986}
\bibinfo{author}{Sellwood, J.}, \bibinfo{author}{Athanassoula, E.},
  \bibinfo{year}{1986}.
\newblock \bibinfo{title}{Unstable modes from galaxy simulations}.
\newblock \bibinfo{journal}{MNRAS} \bibinfo{volume}{221},
  \bibinfo{pages}{195--212}.
\bibitem[{Skokos et~al.(2002)Skokos, Patsis and Athanassoula}]{Skokos}
\bibinfo{author}{Skokos, C.}, \bibinfo{author}{Patsis, P.},
  \bibinfo{author}{Athanassoula, E.}, \bibinfo{year}{2002}.
\newblock \bibinfo{title}{Orbital dynamics of three-dimensional bars--i. the
  backbone of three-dimensional bars. a fiducial case}.
\newblock \bibinfo{journal}{MNRAS} \bibinfo{volume}{333},
  \bibinfo{pages}{847--860}.
\bibitem[{Tremaine and Weinberg(1984)}]{TW1984}
\bibinfo{author}{Tremaine, S.}, \bibinfo{author}{Weinberg, M.D.},
  \bibinfo{year}{1984}.
\newblock \bibinfo{title}{A kinematic method for measuring the pattern speed of
  barred galaxies}.
\newblock \bibinfo{journal}{ApJ} \bibinfo{volume}{282},
  \bibinfo{pages}{L5--L7}.
\bibitem[{Tsoutsis et~al.(2008)Tsoutsis, Efthymiopoulos and
  Voglis}]{Tsoutsis2008}
\bibinfo{author}{Tsoutsis, P.}, \bibinfo{author}{Efthymiopoulos, C.},
  \bibinfo{author}{Voglis, N.}, \bibinfo{year}{2008}.
\newblock \bibinfo{title}{The coalescence of invariant manifolds and the spiral
  structure of barred galaxies}.
\newblock \bibinfo{journal}{Monthly Notices of the Royal Astronomical Society}
  \bibinfo{volume}{387}, \bibinfo{pages}{1264--1280}.
\bibitem[{{Tsoutsis} et~al.(2009){Tsoutsis}, {Kalapotharakos}, {Efthymiopoulos}
  and {Contopoulos}}]{Tsoutsis2009}
\bibinfo{author}{{Tsoutsis}, P.}, \bibinfo{author}{{Kalapotharakos}, C.},
  \bibinfo{author}{{Efthymiopoulos}, C.}, \bibinfo{author}{{Contopoulos}, G.},
  \bibinfo{year}{2009}.
\newblock \bibinfo{title}{{Invariant manifolds and the response of spiral arms
  in barred galaxies}}.
\newblock \bibinfo{journal}{A\&A} \bibinfo{volume}{495},
  \bibinfo{pages}{743--758}.
\bibitem[{Valenzuela and Klypin(2003)}]{Valenz2003}
\bibinfo{author}{Valenzuela, O.}, \bibinfo{author}{Klypin, A.},
  \bibinfo{year}{2003}.
\newblock \bibinfo{title}{Secular bar formation in galaxies with a significant
  amount of dark matter}.
\newblock \bibinfo{journal}{MNRAS} \bibinfo{volume}{345},
  \bibinfo{pages}{406--422}.
\bibitem[{Wu et~al.(2016)Wu, Pfenniger and Taam}]{Wu2016}
\bibinfo{author}{Wu, Y.T.}, \bibinfo{author}{Pfenniger, D.},
  \bibinfo{author}{Taam, R.E.}, \bibinfo{year}{2016}.
\newblock \bibinfo{title}{Time-dependent corotation resonance in barred
  galaxies}.
\newblock \bibinfo{journal}{ApJ} \bibinfo{volume}{830}, \bibinfo{pages}{111}.
\bibitem[{Wu et~al.(2018)Wu, Pfenniger and Taam}]{Wu2018}
\bibinfo{author}{Wu, Y.T.}, \bibinfo{author}{Pfenniger, D.},
  \bibinfo{author}{Taam, R.E.}, \bibinfo{year}{2018}.
\newblock \bibinfo{title}{Time-dependent pattern speeds in barred galaxies}.
\newblock \bibinfo{journal}{ApJ} \bibinfo{volume}{860}, \bibinfo{pages}{152}.

\end{thebibliography}

\end{document}